\documentclass[manuscript]{aastex701}
\usepackage{natbib}
\usepackage[titletoc,title]{appendix}
\usepackage{CJKutf8}
\usepackage{color}
\usepackage{colortbl}
\usepackage{comment}
\usepackage{gensymb}
\usepackage{latexsym, bm}
\usepackage{lineno}
\usepackage{makecell}
\usepackage{multirow}
\usepackage{amsmath}
\usepackage{enumitem}
\usepackage{tikz}
\usetikzlibrary{shapes.geometric, arrows}
\usepackage{txfonts}
\usepackage{url}
\usepackage{xparse}

\ExplSyntaxOn
\NewDocumentCommand{\mylist}{m}
 {
  \seq_set_from_clist:Nn \l_tmpa_seq { #1 }
  \int_set:Nn \l_tmpa_int {\seq_count:N \l_tmpa_seq}
  \int_case:nnF {\l_tmpa_int}
   {
    {1} { \seq_item:Nn \l_tmpa_seq {1} }
    {2} { \seq_item:Nn \l_tmpa_seq {1} \ and\ \seq_item:Nn \l_tmpa_seq {2} }
   }
  {
    \seq_pop_right:NN \l_tmpa_seq \l_tmpb_tl
    \seq_map_function:NN \l_tmpa_seq \__mylist_format:n
    \int_compare:nT {\l_tmpa_int > 1} {\ and\ }
    \l_tmpb_tl
  }
 }
\cs_new:Npn \__mylist_format:n #1 {#1,\ }
\ExplSyntaxOff

\graphicspath{{./}{figures/}}

\usepackage[titletoc,title]{appendix}
\shortauthors{WANG AND QIN}
\shorttitle{coefficients}

\begin{document}
\begin{CJK*}{UTF8}{gbsn}

\newcommand{\qinemail}{qingang@hit.edu.cn}
\newcommand{\gqin}{G. Qin (秦刚)}
\newcommand{\gqinE}{G. Qin}
\newcommand{\gqincode}{\author[0000-0002-3437-3716,gname=Gang,sname=Qin]{\gqin}
}

\newcommand{\wangyangemail}{ywangsz@hit.edu.cn}
\newcommand{\ywang}{Y. Wang (汪洋)}
\newcommand{\ywangE}{Y. Wang}
\newcommand{\ywangcode}{\author[0000-0002-4581-9242]{\ywang}}

\newcommand{\jfwangemail}{wangjunfang@hit.edu.cn}
\newcommand{\jfwang}{J.-F. Wang (王俊芳)}
\newcommand{\jfwangE}{J.-F. Wang}
\newcommand{\jfwangcode}{\author[0000-0002-9586-093X]{\jfwang}}

\newcommand{\lqiaoemail}{liangqiao@hit.edu.cn}
\newcommand{\lqiao}{L. Qiao (乔亮)}
\newcommand{\lqiaoE}{L. Qiao}
\newcommand{\lqiaocode}{\author[0000-0002-4411-2229]{\lqiao}}

\newcommand{\lllianemail}{20231044@jhc.edu.cn}
\newcommand{\lllian}{L.-L. Lian (连乐乐)}
\newcommand{\lllianE}{L.-L. Lian}
\newcommand{\llliancode}{\author[0000-0002-1097-1084]{\lllian}}
\newcommand{\jinhua}{Jinhua University of Vocational
Technology, Jinhua, Zhejiang, 321000, People’s Republic of China}

\newcommand{\sswu}{S.-S. Wu (吴双双)}
\newcommand{\sswuE}{S.-S. Wu}
\newcommand{\sswucode}{\author[0000-0002-5776-455X]{\sswu}}

\newcommand{\swangemail}{wangshu@stu.hit.edu.cn}
\newcommand{\swang}{S. Wang (王姝)}
\newcommand{\swangE}{S. Wang}
\newcommand{\swangcode}{\author[0000-0002-3051-0744]{\swang}}

\newcommand{\xnwangemail}{wangxiaonan@stu.hit.edu.cn}
\newcommand{\xnwang}{X.-N. Wang (王晓男)}
\newcommand{\xnwangE}{X.-N. Wang}
\newcommand{\xnwangcode}{\author[0000-0001-7358-6442]{\xnwang}}

\newcommand{\mldengemail}{dengmeilin@stu.hit.edu.cn}
\newcommand{\mldeng}{M.-L. Deng (邓美林)}
\newcommand{\mldengE}{M.-L. Deng}
\newcommand{\mldengcode}{\author[0000-0002-4576-6152]{\mldeng}}

\newcommand{\yszhongemail}{838693764@qq.com}
\newcommand{\yszhong}{Y.-S. Zhong (仲雨水)}
\newcommand{\yszhongE}{Y.-S. Zhong}
\newcommand{\yszhongcode}{\author[0009-0002-7024-7386]{\yszhong}}

\newcommand{\fjkongemail}{kongfanjing@ncwu.edu.cn}
\newcommand{\fjkong}{F.-J. Kong (孔凡婧)}
\newcommand{\fjkongE}{F.-J. Kong}
\newcommand{\fjkongcode}{\author[0000-0001-7617-8268]{\fjkong}
}

\newcommand{\ncwu}{School of Electronic Engineering, North China University
of Water Resources and Electric Power, Zhengzhou, 450046, People's Republic 
of China}

\newcommand{\aucorres}[1]{\altaffiliation{Author of correspondence.}
\email[show]{#1}
}
\newcommand{\hit}{School of
Science, Harbin Institute of
Technology, Shenzhen, 518055,
People's Republic of China}
\newcommand{\szlab}{Shenzhen Key Laboratory of Numerical Prediction for Space Storm,
Harbin Institute of Technology, Shenzhen, 518055, People's Republic of China}


\newcommand\aastex{AAS\TeX}
\newcommand\latex{La\TeX}

\newcommand{\ttsp}{\hspace{0.25em}}
\newcommand{\tsp}{\hspace{0.5em}}
\newcommand{\wsp}{\hspace{1em}}
\newcommand{\vdag}{(v)^\dagger}
\newcommand{\dee}{\mathrm{d}}
\newcommand{\txt}[1]{\mathrm{#1}}


\newcommand{\asr}{Adv. Space Res.}
\newcommand{\ag}{Ann. Geophys.}
\newcommand{\apr}{Appl. Phys. Res.}
\newcommand{\cpam}{Comm. Pure Appl. Math.}
\newcommand{\cpc}{Comput. Phys. Commun.}
\newcommand{\dossr}{DoSSR}
\newcommand{\jasa}{J. American Statistical Association}
\newcommand{\jastp}{J. Atmos. Sol.-Terr. Phys.}
\newcommand{\jcomph}{J. Comput. Phys.}
\newcommand{\joss}{J. Open Source Software}
\newcommand{\jpcs}{J. Phys.: Conf. Ser.}
\newcommand{\jpg}{J. Phys. G: Nucl. Part. Phys.}
\newcommand{\jplp}{J. Plasma Phys.}
\newcommand{\jswsc}{J. Space Weather Space Clim.}
\newcommand{\lrsp}{Living Rev. Sol. Phys.}
\newcommand{\phpl}{Phys. Plasmas}
\newcommand{\phr}{Phys. Rev.}
\newcommand{\rg}{RvGeo}
\newcommand{\rgsp}{Rev. Geophys. Space Phys.}
\newcommand{\rpp}{Reports Progress Phys.}
\newcommand{\sci}{Science}
\newcommand{\spwea}{Space Weather}
\newcommand{\zg}{Z. Geophys.}
\newcommand{\nnsfc}[1]
{
This work was supported by the grants \mylist{#1}.\hspace{-0.5em}
}

\newcommand{\szstp}[1] 
{
This work was supported by the Shenzhen Science and Technology Program under 
Grant \mylist{#1}.\hspace{-0.5em}
}

\newcommand{\nkrdpc}[1]
{
The authors would like to acknowledge the support of National
Key Research and Development Program of China \mylist{#1}.\hspace{-0.5em}
}

\newcommand{\szkllp}[1]
{
This work was supported by
Shenzhen Key Laboratory Launching Project (\mylist{#1}).\hspace{-0.5em}
}

\newcommand{\sprpcas}[1]
{
This work was supported
by the Strategic Priority Research Program of Chinese Academy of Sciences, 
Grant No. \mylist{#1}.\hspace{-0.5em}
}

\newcommand{\stpguangdong}[1]
{
This work is supported by Science and Technology Program of Guangdong Province 
(grant \mylist{#1}).\hspace{-0.5em}
}

\newcommand{\QinNNSFCelectrons}{NNSFC\ 42074206}
\newcommand{\QinNNSFCouter}{NNSFC\ 42374190}
\newcommand{\QinSZdeeplearning}{No.\ JCYJ20250604145503005}
\newcommand{\QinSZdisaster}{No.\ JCYJ20210324132812029}
\newcommand{\WangjfNNSFC}{NNSFC\ 42374189}
\newcommand{\WangyNNSFC}{NNSFC\ 42474220}
\newcommand{\KongfjNNSFC}{NNSFC\ 42304173}
\newcommand{\GuoNNSFC}{NNSFC\ 42150105}
\newcommand{\GuoNKRDPC}{No.2021YFA0718600}
\newcommand{\ShenNKRDPC}{No.2022YFA1604600}
\newcommand{\ZuoSTPG}{No. 2025B1212050001}
\newcommand{\FengLAB}{No. ZDSYS20210702140800001}

\newcommand{\scTianjin}{The work was carried out at National Supercomputer 
Center in Tianjin, and the calculations were performed on TianHe-3F.
\hspace{0.5em}}
\newcommand{\ace}{ACE}
\newcommand{\acesc}{ACE Science Center}
\newcommand{\epam}{EPAM}

\newcommand{\stereo}{STEREO}
\newcommand{\stereosc}{STEREO Science Centers}

\newcommand{\wind}{Wind}

\newcommand{\psp}{PSP}

\newcommand{\solo}{SolO}

\newcommand{\goes}{GOES}
\newcommand{\xrs}{XRS}

\newcommand{\soho}{SOHO}
\newcommand{\sohodesc}{\soho \hspace{0.5em}is a project
of international cooperation between ESA and NASA.\hspace{0.5em}}
\newcommand{\celias}{CELIAS}

\newcommand{\halphaflare}{``H-alpha Flare” dataset}

\newcommand{\soon}{USAF Solar Observing Optical Network (SOON)}

\newcommand{\noaa}{NOAA}
\newcommand{\ngdc}{NOAA National Geophysical Data Center (NGDC)}

\newcommand{\seon}{Solar Electro-Optical Network (SEON)}
\newcommand{\rstn}{Radio Spectral Telescope Network (RSTN)}

\newcommand{\cmedata}{CME}
\newcommand{\cmelist}{website \url{https://cdaw.gsfc.nasa.gov/CME_list/}}

\newcommand{\cdaweb}{Database for CDAweb (https://cdaweb.gsfc.nasa.gov/)}
\newcommand{\cmecatalog}{The CME catalog is generated and maintained at the 
CDAW Data Center by NASA and the Catholic University of America in
cooperation with the Naval Research Laboratory.\hspace{0.5em}}

\newcommand{\CR}{cosmic ray (CR)}
\newcommand{\nmdb}{Neutron Monitor Database (NMDB)}
\newcommand{\eufp}{European Union's FP7 program (contract No. 213007)}

\newcommand{\scinstr}[2]{#1/#2}

\newcommand{\dataprv}[2]{the #1 data are provided by the #2}
\newcommand{\dataprovideby}[2]{We acknowledge that the \mylist{#1} data are
provided by \mylist{#2}.\hspace{0.5em}}
\newcommand{\dataprovide}[1]{We thank the \mylist{#1} for providing the data 
used in this paper.\hspace{0.5em}}
\newcommand{\datafrommissions}[3]{We acknowledge that the \mylist{#1} data used in
this study are obtained from \mylist{#2}, including observations from the
 \mylist{#3} missions.\hspace{0.5em}}
\newcommand{\dataprovavail}[3]{The #1 are prepared using data provided by the 
#2 and made available through the #3.\hspace{0.5em}}
\newcommand{\dataprovfound}[3]{The #1 data are supplied by the #2, which was 
founded under the #3.\hspace{0.5em}} 
\newcommand{\referee}{The authors thank the anonymous referees for their 
valuable comments.\hspace{0.5em}}

\arraycolsep 0pt

\title{
Moment-based formulas for the transport
coefficients
}


\jfwangcode
\email{jfwangemail}
\affiliation{\hit; \qinemail}

\gqincode
\aucorres{\qinemail}
\affiliation{\hit; \qinemail}
\affiliation{\szlab}

\begin{abstract}

The coefficients of the transport equations
that describe propagation features
are particularly important in astrophysics,
interplanetary physics, and 
experimental plasma physics.	
In this paper, using the momentum transport   
equations, we find that	the coefficients 
are related to 
statistical quantities, some of these relations
are new. 
In addition, these coefficients 
take logarithmic forms that 
differ from those derived
in previous studies. Moreover, we investigate
variable-dependent diffusion 
coefficients in transport equations.
For isotropic pitch-angle scattering,
we also obtain a formula that has 
a logarithmic form. 
For fractional transport
equations, 
the coefficients expressed in terms of the moments
are obtained from the governing equations.

\end{abstract}


\section{INTRODUCTION}

The transport of energetic charged particles 
in turbulent magnetic fields of plasmas
has been widely recognized as important 
in space plasma physics and experimental plasmas
\citep{Jokipii1966, MatthaeusEA2003, WebbEA2006, 
Shalchi2009, Zank2014, Shalchi2020}.
Of particular
interest are transport coefficients 
for the spatial, momentum, and pitch-angle 
scattering, which control the penetration and modulation of cosmic rays in
the heliosphere, and
the efficiency of diﬀusive shock 
acceleration 
\citep{Jokipii1966,  Schlickeiser2002, MatthaeusEA2003, Qin2007, 
Shalchi2010, Shalchi2020}. 
So far, the transport coefficients
have two different expressions,
i.e., Taylor-Green-Kubo formulation 
(hereafter Kubo formula)
and formulas in terms of
the moments (hereafter FIM)
\citep{Taylor1922, Green1951,
Kubo1962}.   
Based on the Kubo formula, various perpendicular and parallel models have been developed,
the BAM model \citep{BieberMatthaeus1997},
the nonlinear guiding center theory (NLGC)
\citep{MatthaeusEA2003}, 
the nonlinear parallel diffusion theory
\citep{Qin2007},
the unified nonlinear theory
\citep{Shalchi2010}, 
the modification of NLGC
\citep{QinEA2014}, and others. 
However, in studies of
spatial transport coefficients
with adiabatic focusing
(i.e., a spatially non-uniform background 
magnetic field),
scientists have found that 
the Kubo formula and FIM yield different results
\citep{Roelof1969,
Earl1976,Kunstmann1979,BeeckEA1986,BieberEA1990,
Kota2000, SchlickeiserEA2008, Shalchi2011,
Litvinenko2012a,Litvinenko2012b,
DanosEA2013,ShalchiEA2013,HeEA2014,WangEA2016,
WangEA2017b}.
\citet{wq2018, wq2019, wq2020} and
demonstrated
that FIM
has a broader range of applicability. 
Thereafter, the relationship between FIM
and the moments has been thoroughly
investigated\citep{wq2020}.

For the linear spatial transport equation with
constant coefficients,
\citet{wq2023} derived formulas for the 
coefficients of  
for first, second, and higher-order 
derivative terms. 
They found 
that these coefficients are all equal to
the time derivative of 
statistical quantities
that are functions of the moments.
For example, the coefficient 
of the convection equation
is the time derivative of the expectation;
the coefficient of
the diffusion equation
is the time derivative of the variance;
and the coefficients of higher-order transport
equations are all the time derivatives of different
statistical quantities
\citep{wq2023}. 
When we define a statistical quantity as 
$Q$ and a constant transport coefficient as $\kappa$, 
the results obtained by \citet{wq2023}
can be expressed in a unified form 
at late times either as
\begin{eqnarray}
\kappa=\frac{\dee}{\dee{t}}Q,
\end{eqnarray}
or as
\begin{eqnarray}
Q=\kappa t.
\end{eqnarray}
The latter equation
indicates that a statistical quantity
is a linear function of time at late times.
These results were derived from the spatial
transport equations. 
However, the relationship
between momentum transport coefficients and
momentum moments has not been explored. 
In this paper, we investigate the
relationship between momentum transport 
coefficients and momentum moments. 
In addition, in general,  
the coefficients of transport equations
are not constant and depend on 
time and position. 
Therefore, we need to explore 
the characteristics of 
variable transport coefficients.

In this article, 
we investigate the variable 
transport coefficients.
This article is organised as follows. 
In Section 
\ref{Moment-based momentum transport coefficients}, 
the momentum transport equations is derived from
the Fokker-Planck equation with variable
coefficients. 
The logarithmic relationship between the coefficients
and statistical quantities is observed in our
calculations.
In addition, some new statistical quantities
are derived to express the momentum
transport coefficients.
In Section \ref{Moment-based pitch-angle diffusion coefficient}, 
the pitch-angle scattering coefficient
taking a logarithmic form is derived.
In Section 
\ref{Moment-based the Fokker-Planck coefficient},
we derive the coefficients of the 
variable coefficient Fokker-Planck equations
and again obtain the logarithmic form.
In addition, some well-known Fokker-Planck-type
equations with variable coefficients
in statistical mechanics 
are investigated, and the corresponding coefficient formulas
are deduced.
In Section 
\ref{Moment-based Fractional transport equation 
	coefficient},
the coefficients of fractional equations 
are explored.
The results are presented 
in Section 
\ref{SUMMARY AND CONCLUSION}.

\section{Moment-based formulas for momentum transport coefficients}
\label{Moment-based momentum transport coefficients}
In this section, we explore the momentum transport
equations and derive
the formulas for the coefficients in terms of
statistical quantities derived from
the variable moments.

\subsection{Transport equation for 
energetic particles in momentum space 
}
\label{The energetic-particle momentum 
	transport equation}

The transport of energetic cosmic-ray particles 
in turbulent 
magnetized plasmas, such as the interplanetary 
and interstellar 
media, can be described by the following 
Fokker-Planck equation
\begin{eqnarray}
\frac{\partial{f}}{\partial{t}}
=-\nabla\cdot\left(\kappa_r f\right)
-\nabla_{\bm{p}}\cdot\left(\kappa_p f\right)
+\nabla_{\bm{p}}^2\cdot
\left(\kappa_{pp} f\right).	
\label{Fokker-Planck equation}	
\end{eqnarray}
Here, $f=f(\bm{r}, \bm{p}, t)$ is 
the particle distribution function
with position $\bm{r}$
and momentum $\bm{p}$, 
$\kappa_r$ is the 
spatial convection coefficient, 
$\kappa_p$ is the 
momentum convection coefficient, 
and $\kappa_{pp}$ is the 
momentum diffusion coefficient. 
After averaging over perpendicular spatial 
coordinates and the gyrophase in 
spherical momentum coordinates,  
we obtain from 
Equation 
(\ref{Fokker-Planck equation})
\begin{eqnarray}
\frac{\partial{f}}{\partial{t}}
&=&
-v\mu\frac{\partial{f}}{\partial{z}}
-\frac{1}{p^2}\frac{\partial{}}{\partial{p}}
(p^2 D_{p}f)
+\frac{\partial{}}{\partial{\mu}}
\left(D_{\mu\mu}
\frac{\partial{f}}{\partial{\mu}}
+D_{\mu p}
\frac{\partial{f}}{\partial{p}}
\right)
\nonumber\\
&&
+\frac{1}{p^2}\frac{\partial{}}{\partial{p}}
\left(p^2 D_{pp}\frac{\partial{f}}{\partial{p}}
+p^2 D_{p\mu}\frac{\partial{f}}
{\partial{\mu}}
\right).
\label{original Fokker-Planck equation}				
\end{eqnarray}
This equation contains spatial convection, 
momentum convection,
momentum diffusion, pitch-angle diffusion,
and mixed-derivative terms with coefficients $D_{\mu p}$ or $D_{p\mu}$. Here, the pitch-angle 
convection is neglected.  
To explore whether any new results could 
be obtained and to simplify the derivation, 
we use the following simpler model as our 
starting point
\begin{eqnarray}
\frac{\partial{f}}{\partial{t}}
&=&
-v\mu\frac{\partial{f}}{\partial{z}}
-\frac{1}{p^2}\frac{\partial{}}{\partial{p}}
(p^2 D_{p}f)
+\frac{\partial{}}{\partial{\mu}}
\left(D_{\mu\mu}
\frac{\partial{f}}{\partial{\mu}}\right)
+\frac{1}{p^2}\frac{\partial{}}{\partial{p}}
(p^2 D_{pp}\frac{\partial{f}}{\partial{p}}),	
\label{Fokker-Planck equation used}	
\end{eqnarray}
with the momentum convection and diffusion coefficients,
and pitch-angle scattering coefficient as 
\begin{eqnarray}
&&D_{\mu\mu}=D_{\mu\mu}(\mu), \\	
&&D_{p}=p A(\mu), \\
&&D_{pp}=N p^2D_{\mu\mu}(\mu).
\label{Fokker-Planck momentum coefficients}			
\end{eqnarray}
Here, $f=f(z, p, \mu, t)$,
$A(\mu)$ is a function of $\mu$,
and $N$ is constant
\citep{Schlickeiser2002}. 
In Equation
(\ref{Fokker-Planck equation used}),
the  mixed-derivative terms with
$D_{\mu p}$ or $D_{p\mu}$ are neglected.
A more detailed study of 
Equation 
(\ref{original Fokker-Planck equation})
will be left for future work. 
In this paper, $D_{\mu\mu}$ is used 
frequently to derive the momentum transport 
equation from the Fokker-Planck equation, 
as shown in Appendix 
\ref{The derivation of the momentum transport equation}.
In order to simplify the derivation, 
a simpler form of $D_{\mu\mu}$ is used;
that is, we assume $D_{\mu\mu}=D(1-\mu^2)$
with constant $D$. 
Simplified forms of $D_{p}$ and $D_{pp}$ 
are also employed in this paper. 
In future work, 
the coefficient formulas will be explored
for cases in which 
$D_{\mu\mu}$, $D_{p}$ and $D_{pp}$ 
depend on other variables
or take other forms. 	     

The distribution function $f(z, p, \mu, t)$ in
Equation 
(\ref{Fokker-Planck equation used})
can be decomposed into the 
isotropic part $F(z, p, t)$ and the anisotropic part 
$g(z, p, \mu, t)$ 
\begin{eqnarray}
f(z, p, \mu, t)=F(z, p, t)+g(z, p, \mu, t).
\label{f=F+g}
\end{eqnarray}
Inserting the latter formula into Equation 
(\ref{Fokker-Planck equation used}) and integrating
over $z$ gives (the detailed derivation is shown in 
Appendix 
\ref{The derivation of the momentum transport equation})
\begin{eqnarray}
\frac{\partial{F}}{\partial{t}}
&=&-\mathcal{K}_p
\frac{1}{p^2}\frac{\partial{}}{\partial{p}}
\left(p^{3}F\right)
+
\mathcal{K}_1
\frac{1}{p^2}\frac{\partial{}}{\partial{p}}
\left(p^{4}
\frac{\partial{F}}{\partial{p}}\right)	
+\mathcal{K}_2
\frac{1}{p^2}\frac{\partial{}}{\partial{p}}
\left[p	
\frac{\partial{}}{\partial{p}}
\left(p^{3}F\right)\right].
\label{momentum transport equation as starting point}			
\end{eqnarray}
Here, the first term on the right-hand side 
represents the momentum convection,
with $\mathcal{K}_p$ as its 
coefficient;
the second term represents the first 
momentum diffusion, with $\mathcal{K}_1$ as its 
coefficient;
the third term represents the second
momentum diffusion, with $\mathcal{K}_2$ 
as its coefficient. 

In this work, the following formula for the
$i$th moment of momentum is frequently used 
\begin{eqnarray}
\left\langle p^i\right\rangle=
\int_{-\infty}^{+\infty}p^i F(p,t)
\dee p_x 
\dee p_y \dee p_z
=4\pi\int_0^{+\infty} p^i F(p,t)p^2\dee p,	
\label{momentum moment}
\end{eqnarray}
where, $i$ is a natural numbers. 
Below,
we derive the moment-based transport
coefficients for the special case
of Equation 
(\ref{momentum transport equation as starting point}). 
    
\subsection{Moment-based formula for 
momentum convection coefficients} 
\label{Moment-based momentum convection coefficients}

Here, we derive the transport coefficient 
for the following momentum convection equation
\begin{eqnarray}
\frac{\partial{F}}{\partial{t}}
&=&-\mathcal{K}_p
\frac{1}{p^2}\frac{\partial{}}{\partial{p}}
\left(p^{3}F\right),	
\label{momentum convection equation}
\end{eqnarray}
which contains only the first term on the 
right-hand side of Equation
(\ref{momentum transport equation as starting point}). 
Integrating Equation 
(\ref{momentum convection equation})
by parts, we obtain
the first-order moment equation for 
the momentum distribution 
\begin{eqnarray}
\frac{\dee }{\dee t}\left\langle p\right\rangle
&=&\mathcal{K}_p \left\langle p\right\rangle.	
\label{first momentum of momentum from
momentum convection equation}							
\end{eqnarray}
The solution of Equation
(\ref{first momentum of momentum from
momentum convection equation})
is 
\begin{eqnarray}
\left\langle p\right\rangle=\left\langle p\right\rangle_0 e^{\mathcal{K}_pt},
\label{solution}
\end{eqnarray}
where $\left\langle p\right\rangle_0$ is the initial
value of $\left\langle p\right\rangle$.
From Equation 
(\ref{solution}), we obtain
\begin{eqnarray}
\mathcal{K}_p=\frac{\dee }{\dee t}
\left[\ln
\left\langle p\right\rangle
-\left\langle p\right\rangle_0
\right]
=
\frac{\dee }{\dee t}
\ln
\frac{\left\langle p\right\rangle}
{\left\langle p\right\rangle_0}
=\frac{\dee }{\dee t}
\ln
\left\langle p\right\rangle
,	
\label{convection coefficient}								
\end{eqnarray}
which implies that the momentum coefficient
can be expressed in terms of 
the logarithm of the first-order
moment. 
In fact, from Equation
(\ref{first momentum of momentum from
	momentum convection equation}),
we can also obtain Equation
(\ref{convection coefficient}).
If we define the first moment as
\begin{eqnarray} 
Q_1=\left\langle p\right\rangle,
\label{the new statistical quantity
	with logarithmic form}	
\end{eqnarray}
the momentum convection 
coefficient can be rewritten as 
\begin{eqnarray}
\mathcal{K}_p 
=
\frac{\dee}{\dee t}\ln Q_1.							
\end{eqnarray}
The latter equation shows 
that 
the momentum convection coefficient
is a function of $Q_1$.
In a previous study, 
\citet{wq2023} found that, 
for constant-coefficient spatial transport equations,
the coefficients depend on
statistical quantities. 
Here, we find the similar result.
However, for momentum transport, the convection
coefficient takes a logarithmic form.

\subsection{Moment-based formula for 
momentum diffusion coefficients}
\label{Momentum diffusion coefficients in terms 
of moments}

If we neglect the convection term
in Equation
(\ref{momentum transport equation as starting point}), the momentum transport equation,
which contains the first and second diffusion terms,
can be written as 
\begin{eqnarray}
\frac{\partial{F}}{\partial{t}}
&=&
\mathcal{K}_1
\frac{1}{p^2}\frac{\partial{}}{\partial{p}}
\left(p^{4}
\frac{\partial{F}}{\partial{p}}\right)	
+\mathcal{K}_2
\frac{1}{p^2}\frac{\partial{}}{\partial{p}}
\left[p	
\frac{\partial{}}{\partial{p}}
\left(p^{3}F\right)\right].
\label{momentum diffusion equation with two terms}				
\end{eqnarray}
In what follows, we investigate the 
relationship between momentum diffusion coefficients
and statistical quantities.

\subsubsection{
The case 1. }
\label{The momentum diffusion equation with the first diffusion term}
Here, we consider 
the momentum diffusion equation  with
only the first diffusion term,   
\begin{eqnarray}
\frac{\partial{F}}{\partial{t}}
&=&
\mathcal{K}_1
\frac{1}{p^2}\frac{\partial{}}{\partial{p}}
\left(p^{4}
\frac{\partial{F}}{\partial{p}}\right),
\label{The momentum diffusion equation I.}		
\end{eqnarray}
and derive the formula for the 
momentum diffusion coefficient 
$\mathcal{K}_1$.
With formula (\ref{momentum moment}),
the first order moment 
can be derived from the latter equation
\begin{eqnarray}
\frac{\dee}{\dee{t}}\left\langle p\right\rangle
&=&4
\mathcal{K}_1
\left\langle p\right\rangle,						
\end{eqnarray}
and the corresponding momentum diffusion coefficient
can be found 
\begin{eqnarray}
\mathcal{K}_1
=
\frac{\dee}{\dee{t}}
\ln\left(
\left\langle p\right\rangle^{1/4}\right).	
\label{momentum diffusion coefficient I for i=1}			
\end{eqnarray}
By setting 
\begin{eqnarray}  
Q_2=
\left\langle p\right\rangle^{1/4},
\label{Q2}
\end{eqnarray}
we can rewrite 
Equation 
(\ref{momentum diffusion coefficient I for i=1})
as
\begin{eqnarray}
\mathcal{K}_1
=
\frac{\dee}{\dee{t}}
\ln Q_2.	
\label{momentum diffusion coefficient I with Q_2}				
\end{eqnarray}
This formula indicates that 
the first momentum diffusion coefficient 
is related to the fourth root 
of the first moment.

\subsubsection{The Case 2.}
\label{The momentum diffusion equation with
the second diffusion term}
	
Retaining only the second term in Equation
(\ref{momentum diffusion equation with two terms}),
we obtain
\begin{eqnarray}
\frac{\partial{F}}{\partial{t}}
&=&	
\mathcal{K}_2
\frac{1}{p^2}\frac{\partial{}}{\partial{p}}
\left[p	
\frac{\partial{}}{\partial{p}}
\left(p^{3}F\right)\right].
\label{momentum diffusion equation II}				
\end{eqnarray}
From the latter equation, we can obtain the first order
moment equation of momentum as
\begin{eqnarray}
\frac{\dee}{\dee{t}}\left\langle p\right\rangle
=\mathcal{K}_2
\left\langle p\right\rangle.
\label{first order moment equation for diffusion equation II}					
\end{eqnarray}
Employing this equation, we can derive 
the formula for 
the momentum diffusion
coefficient
\begin{eqnarray}
\mathcal{K}_2
=
\frac{\dee}{\dee{t}}
\ln
\left\langle p\right\rangle.
\end{eqnarray}
With Equation 
(\ref{the new statistical quantity
with logarithmic form}), the latter equation 
can be rewritten as
\begin{eqnarray}
\mathcal{K}_2
=
\frac{\dee}{\dee{t}}\ln
Q_1.							
\end{eqnarray}
This formula 
shows that the second momentum diffusion
coefficient $\mathcal{K}_2$  
takes the same form as the momentum 
convection coefficient $\mathcal{K}_p$. 
They are all determined by the 
first-order moment $Q_1$.

\subsubsection{The Case 3.}
\label{The momentum diffusion equation 
	with the first and second diffusion terms}

Here, we investigate the momentum diffusion coefficients
for the following equation
\begin{eqnarray}
\frac{\partial{F}}{\partial{t}}
&=&
\mathcal{K}_1
\frac{1}{p^2}\frac{\partial{}}{\partial{p}}
\left(p^{4}
\frac{\partial{F}}{\partial{p}}\right)	
+\mathcal{K}_2
\frac{1}{p^2}\frac{\partial{}}{\partial{p}}
\left[p	
\frac{\partial{}}{\partial{p}}
\left(p^{3}F\right)\right],
\label{momentum transport equation with 
I and II}				
\end{eqnarray}
which contains two different momentum diffusion terms.
The first-order moment equation for the momentum 
distribution 
can be readily derived 
\begin{eqnarray}
\frac{\dee}{\dee{t}}\left\langle p\right\rangle
&=&4\mathcal{K}_1
\left\langle p\right\rangle
+\mathcal{K}_2
\left\langle p\right\rangle,	
\end{eqnarray}
which can be rewritten as
\begin{eqnarray}
\frac{\dee}{\dee{t}}
\ln
\left\langle p\right\rangle
&=&4\mathcal{K}_1
+\mathcal{K}_2.
\label{first order momentum moment equation for two diffusion term}							
\end{eqnarray}
The latter equation is linear in
$\mathcal{K}_1$ and $\mathcal{K}_2$.
In order to derive the formulas
for $\mathcal{K}_1$ and $\mathcal{K}_2$,
we need to obtain another equation for the momentum
distribution, 
e.g., 
the second-order moment equation,
\begin{eqnarray}
\frac{\dee}{\dee{t}}
\ln
\left\langle p^2\right\rangle
&=&10
\mathcal{K}_1
+4\mathcal{K}_2.	
\label{second order momentum moment equation for two diffusion term}								
\end{eqnarray}
Combining Equations
(\ref{first order momentum moment equation for two diffusion term})
and 
(\ref{second order momentum moment equation for two diffusion term})
yields
\begin{eqnarray}
&&\mathcal{K}_1
=
\frac{\dee}{\dee{t}}
\ln
\left(
\frac{\left\langle p\right\rangle^4}
{\left\langle p^2\right\rangle}
\right)^{1/6},
\label{k1 formula for two diffusion}
\\	
&&\mathcal{K}_2
=
\frac{\dee}{\dee{t}}
\ln
\left(
\frac{\left\langle p^2\right\rangle^{2}}
{\left\langle p\right\rangle^5}
\right)^{1/3}.
\label{k2 formula for two diffusion}
\end{eqnarray}
Defining two statistical quantities as
\begin{eqnarray}
&&Q_3=
\left(
\frac{\left\langle p\right\rangle^4}
{\left\langle p^2\right\rangle}
\right)^{1/6},\\
&&Q_4=
\left(
\frac{\left\langle p^2\right\rangle^{2}}
{\left\langle p\right\rangle^5}
\right)^{1/3}, 
\end{eqnarray}
we rewrite Equations 
(\ref{k1 formula for two diffusion})
and
(\ref{k2 formula for two diffusion})
as
\begin{eqnarray}
&&\mathcal{K}_1
=
\frac{\dee}{\dee{t}}\ln
Q_3,
\label{k1 formula for two diffusion with Q3}
\\	
&&\mathcal{K}_2
=
\frac{\dee}{\dee{t}}\ln
Q_4.
\label{k2 formula for two diffusion with Q4}	
\end{eqnarray}
Here, $Q_3$ and $Q_4$ are two new statistical 
quantities.

\subsection{Moment-based formulas for 
the momentum transport coefficients}
\label{Moment-based momentum equation transport coefficients 
with convection and diffusion ones}

In this subsection, we explore the momentum 
transport equation with the convection term
and one of the diffusion terms. 

\subsubsection{The case 1.}
\label{The case 1.}

Here, we investigate the transport
coefficients in the momentum equation
with the convection term and the first diffusion
term 
\begin{eqnarray}
\frac{\partial{F}}{\partial{t}}
=-\mathcal{K}_p
\frac{1}{p^2}\frac{\partial{}}{\partial{p}}
\left(p^{3}F\right)
+
\mathcal{K}_1
\frac{1}{p^2}\frac{\partial{}}{\partial{p}}
\left(p^{4}
\frac{\partial{F}}{\partial{p}}\right).
\label{the first convection-diffusion equation}				
\end{eqnarray}
From the convection-diffusion equation for 
the momentum distribution,
we derive the first-order moment equation
\begin{eqnarray}
\frac{\dee}{\dee{t}}\left\langle p\right\rangle
=
\mathcal{K}_p
\left\langle p\right\rangle
+4\mathcal{K}_1
\left\langle p\right\rangle,	
\end{eqnarray}
which can be rewritten as
\begin{eqnarray}
\frac{\dee}{\dee{t}}
\ln
\left\langle p\right\rangle
=
\mathcal{K}_p
+4\mathcal{K}_1.	
\label{first-order moment equation for the first
convection-diffusion equation}					
\end{eqnarray}
Similarly, we derive the equation for the 
second-order moment of the momentum distribution
\begin{eqnarray}
\frac{\dee}{\dee{t}}
\ln
\left\langle p^2\right\rangle
=2
\mathcal{K}_p
+10\mathcal{K}_1.	
\label{second-order moment equation for the first
convection-diffusion equation}					
\end{eqnarray}
Considering Equations 
(\ref{first-order moment equation for the first
convection-diffusion equation})
and
(\ref{second-order moment equation for the first
convection-diffusion equation})
gives
\begin{eqnarray}
&&\mathcal{K}_p=\frac{\dee}{\dee{t}}
\ln
\frac{\left\langle p\right\rangle^5}
{\left\langle p^2\right\rangle^2},
\label{kp for the first convection-diffusion 
equation}\\
&&\mathcal{K}_1=
\frac{\dee}{\dee{t}}
\ln
\left(
\frac{\left\langle p^2\right\rangle}
{\left\langle p\right\rangle^2}
\right)^{1/2}.
\label{k1 for the first convection-diffusion 
equation}
\end{eqnarray}
Defining the statistical quantities
\begin{eqnarray}
&&Q_5=\frac{\left\langle p\right\rangle^5}
{\left\langle p^2\right\rangle^2},\\
&&Q_6=\left(
\frac{\left\langle p^2\right\rangle}
{\left\langle p\right\rangle^2}
\right)^{1/2},
\label{Q6}
\end{eqnarray}
we rewrite Equations
(\ref{kp for the first convection-diffusion 
equation})
and 
(\ref{k1 for the first convection-diffusion 
equation})
as
\begin{eqnarray}
&&\mathcal{K}_p=\frac{\dee}{\dee{t}}
\ln
Q_5,
\label{kp for the first convection-diffusion 
	equation}\\
&&\mathcal{K}_1=
\frac{\dee}{\dee{t}}
\ln
Q_6.
\end{eqnarray}
The statistical quantities $Q_5$ and $Q_6$
are new. 

\subsubsection{The case 2.}
\label{convection and second diffusion}

Retaining the convection term and the second
diffusion term in Equation
(\ref{momentum transport equation as starting point}),
yields the  
following momentum convection-diffusion equation
\begin{eqnarray}
\frac{\partial{F}}{\partial{t}}
&=&-\mathcal{K}_p
\frac{1}{p^2}\frac{\partial{}}{\partial{p}}
\left(p^{3}F\right)
+\mathcal{K}_2
\frac{1}{p^2}\frac{\partial{}}{\partial{p}}
\left[p	
\frac{\partial{}}{\partial{p}}
\left(p^{3}F\right)\right].
\label{The case 2.}				
\end{eqnarray}
From the latter equation, 
the first- and second-order moment equations 
for the momentum distribution
are derived as
\begin{eqnarray}
&&\frac{\dee}{\dee{t}}
\ln
\left\langle p\right\rangle
=
\mathcal{K}_p
+
\mathcal{K}_2,	
\label{the first-order moment equation for 
the second convection-diffusion equation}\\
&&\frac{\dee}{\dee{t}}
\ln
\left\langle p^2\right\rangle
=2
\mathcal{K}_p
+
4\mathcal{K}_2.
\label{the second-order moment equation for 
the second convection-diffusion equation}						
\end{eqnarray}
From the latter two equations, we derive
\begin{eqnarray}
&&\mathcal{K}_p=\frac{\dee}{\dee{t}}
\ln
\left(
\frac{\left\langle p\right\rangle^4}
{\left\langle p^2\right\rangle}
\right)^{1/2},
\label{kp for the first convection-diffusion 
	equation}\\
&&\mathcal{K}_2=
\frac{\dee}{\dee{t}}
\ln
\left(
\frac{\left\langle p^2\right\rangle}
{\left\langle p\right\rangle^2}
\right)^{1/2}.
\label{k1 for the first convection-diffusion 
	equation}	
\end{eqnarray}
Defining the following statistical quantity
\begin{eqnarray}
Q_7=\left(
\frac{\left\langle p\right\rangle^4}
{\left\langle p^2\right\rangle}
\right)^{1/2}
\end{eqnarray}
and using the formula for statistical quantity
$Q_6$, we rewrite Equations
(\ref{kp for the first convection-diffusion 
	equation}) 
and 
(\ref{k1 for the first convection-diffusion 
	equation})
as
\begin{eqnarray}
&&\mathcal{K}_p=\frac{\dee}{\dee{t}}
\ln
Q_7,
\label{kp for the second convection-diffusion 
	equation}\\
&&\mathcal{K}_2=
\frac{\dee}{\dee{t}}
\ln
Q_6.
\label{k2 for the second convection-diffusion 
	equation}		
\end{eqnarray}
Here, $Q_7$ is a new statistical quantity.  

\section{Moment-based formula for 
pitch-angle diffusion coefficient}
\label{Moment-based pitch-angle diffusion coefficient}

The pitch-angle scattering 
of energetic particles leads to 
parallel spatial diffusion via isotropization. 
This process is described by the pitch-angle
diffusion coefficient $D_{\mu\mu}$, 
whose formula we derive in this section. 
The pitch-angle scattering equation is given by
\begin{eqnarray}
\frac{\partial{F}}{\partial{t}}
&=&	
\frac{\partial{}}{\partial{\mu}}
\left(D_{\mu\mu}	
\frac{\partial{F}}{\partial{\mu}}
\right),				
\end{eqnarray}
where $F=F(\mu, \mu_0, t)$.
For strong magnetic turbulence,
the formula takes the form \citep{ShalchiEA2009}
\begin{eqnarray}
D_{\mu\mu}=D(1-\mu^2)
\label{isotropic pitch-angle diffusion coefficient}
\end{eqnarray}
with constant $D$. 
Thus, the pitch-angle scattering
equation can be rewritten as
\begin{eqnarray}
\frac{\partial{F}}{\partial{t}}
&=&	
D\frac{\partial{}}{\partial{\mu}}
\left((1-\mu^2)	
\frac{\partial{F}}{\partial{\mu}}
\right).	
\label{pitch-angle scattering equation}					
\end{eqnarray}
The pitch-angle correlation is given by
\begin{eqnarray}
\left\langle \mu\mu_0\right\rangle
=
\int_{-1}^{+1}\dee\mu_0
\int_{-1}^{+1}\dee\mu \mu\mu_0 F(\mu, \mu_0, t),	
\end{eqnarray}
where $\mu$ is pitch-angle, 
$\mu_0$ is initial pitch-angle, 
and $t$ is the time.  
From Equation
(\ref{pitch-angle scattering equation}),
we find 
\begin{eqnarray}
D=-\frac{1}{2}
\frac{\dee}{\dee{t}}
\ln 
\left\langle \mu\mu_0\right\rangle. 					 				
\end{eqnarray}
With formula 
(\ref{isotropic pitch-angle diffusion coefficient}),
the latter equation can be rewritten as
\begin{eqnarray}
D_{\mu\mu}=-\frac{1}{2}
\left(1-\mu^2\right)
\frac{\dee}{\dee{t}}
\ln 
\left\langle \mu\mu_0\right\rangle,	 					 				
\end{eqnarray}
which is identical to the result obtained
by \citet{Tautz2013}.
Similarly, we can derive the second-order
moment of the pitch-angle distribution as 
\begin{eqnarray}
D=-\frac{1}{6}\frac{\dee}{\dee{t}}
\ln \left(1-3
\left\langle \mu^2\right\rangle
\right).			 				
\end{eqnarray}
Using Equation 
(\ref{isotropic pitch-angle diffusion coefficient}),
we obtain the following formula
\begin{eqnarray}
D_{\mu\mu}=-\frac{1}{6}
\left(1-\mu^2\right)
\frac{\dee}{\dee{t}}
\ln 
\left(1-3
\left\langle \mu^2\right\rangle
\right).				 				
\end{eqnarray}

\section{Moment-based formulas for 
Fokker-Planck coefficients}
\label{Moment-based the Fokker-Planck coefficient}

In this section, we investigate 
the coefficients of the Fokker-Planck equation, 
which takes the form
\begin{eqnarray}
\frac{\partial{F}}{\partial{t}}
=
-\frac{\partial{}}{\partial{x}}	
\left[A(x)F(x,t)\right]
+
\frac{\partial^2{}}{\partial{x^2}}	
\left[B(x)F(x,t)\right],			 				
\end{eqnarray}
where $A(x)$ and $B(x)$ are the 
first- and second-order coefficients, respectively.

\subsection{Moment-based formulas for the first-order Fokker-Planck coefficient}
\label{The first-order Fokker-Planck equation}

Here, we explore the coefficient of the first-order
Fokker-Planck equation
\begin{eqnarray}
\frac{\partial{F}}{\partial{t}}
=
-\frac{\partial{}}{\partial{x}}	
\left[A(x)F(x,t)\right],
\label{first-order Fokker-Planck equation}			 				
\end{eqnarray}	
Using the formula for the first-order moment 
\begin{eqnarray}
\left\langle x\right\rangle
=\int x F(x, t)\dee x, 
\label{moment formula}
\end{eqnarray}
we derive from Equation 
(\ref{first-order Fokker-Planck equation})
\begin{eqnarray}
\frac{\dee}{\dee t}\left\langle x\right\rangle
=
\left\langle A(x)\right\rangle.
\label{first-order moment equation}
\end{eqnarray}
The latter formula indicates 
that the ensemble average 
of the variable coefficient 
is a function of the expectation of $x$. 
When $A(x)=\kappa_x$ is constant, 
Equation
(\ref{first-order moment equation})
becomes
\begin{eqnarray}
\kappa_x
=
\frac{\dee}{\dee t}
\left\langle x\right\rangle.	
\end{eqnarray}
If the first-order coefficient is a linear 
function of $x$, i.e., 
$A(x)=\kappa_x x$ with constant $\kappa_x$, 
we obtain
\begin{eqnarray}
\kappa_x
=
\frac{\dee}{\dee t}
\ln
\left\langle x\right\rangle.			
\end{eqnarray}
We find that the coefficient
$\kappa_x$ takes a logarithmic form
in this case.
In fact, the first-order Fokker-Planck
coefficient can take a more general form, 
e.g., $A(x)=\kappa_x x^n$
where $\kappa_x$ is constant. 
However, for this
general form, we find 
\begin{eqnarray}
\kappa_x
=\frac{1}{n}
\frac{1}{\left\langle x^{n-1}\right\rangle}
\frac{\dee}{\dee t}
\left\langle x^n\right\rangle,		
\label{coefficient with variable time}		
\end{eqnarray}
which cannot be written as the time derivative 
of 
a single statistical quantity. 
Since the right-hand side of the following
formula depends on time $t$
\begin{eqnarray}
\left\langle x^{n-1}\right\rangle
=\int x^{n-1} F(x, t)\dee x,
\label{moment with time}
\end{eqnarray}
the moment 
$\left\langle x^{n-1}\right\rangle$
is a function of time. 
Substituting formula
(\ref{moment with time}) into
(\ref{coefficient with variable time}),
we find that the coefficient $\kappa_x$
is a function of time,
i.e., 
$\kappa_x=\kappa_x(t)$. However, 
$\kappa_x=\kappa_x(t)$ contradicts 
the assumption that 
$\kappa_x$ is constant.  
If we set $A(x, t)=\kappa_x(t) x$,
we can rewrite Equation
(\ref{first-order Fokker-Planck equation}) as
\begin{eqnarray}
\frac{\partial{F}}{\partial{t}}
=
-\frac{\partial{}}{\partial{x}}	
\left[\kappa_x(t) xF(x,t)\right].
\label{first-order Fokker-Planck equation with k(t)}				 				
\end{eqnarray}	 
From this equation, the formula for 
the coefficient can be written as
\begin{eqnarray}
\kappa_x(t)
=\frac{1}{n}
\frac{1}{\left\langle x^{n-1}\right\rangle}
\frac{\dee}{\dee t}
\left\langle x^n\right\rangle.		
\end{eqnarray}
We will explore the time-dependent 
coefficient in future work.

\subsection{Moment-based formulas for the second-order Fokker-Planck coefficient}
\label{The second-order equation}

In this subsection, we investigate 
the coefficient of the 
second-order Fokker-Planck equation 
\begin{eqnarray}
\frac{\partial{F}}{\partial{t}}
=
\frac{\partial^2{}}{\partial{x^2}}	
\left[B(x)F(x,t)\right].			 				
\end{eqnarray}
Here, $B(x)$ is the second-order 
Fokker-Planck coefficient. 
Using Equation
(\ref{moment formula}),
we can obtain the equations for
the first- and second-order 
moments as
\begin{eqnarray}
&&\frac{\dee}{\dee t}
\left\langle x\right\rangle
=0,\\
&&\frac{\dee}{\dee t}
\left\langle x^2\right\rangle
=
2\left\langle B(x)\right\rangle.		
\end{eqnarray}
If the second-order Fokker-Planck
coefficient is constant, i.e., with
$B(x)=\kappa_{xx}$ constant,
we find 
\begin{eqnarray}
\kappa_{xx}
=
\frac{1}{2}
\frac{\dee}{\dee t}
\left\langle x^2\right\rangle.				
\end{eqnarray}
If the second-order Fokker-Planck
coefficient is quadratic in 
$x$, i.e., 
$B(x)=\kappa_{xx}x^2$ with $\kappa_{xx}$
constant,
we derive
\begin{eqnarray}
\kappa_{xx}
=
\frac{\dee}{\dee t}\ln\left(
\left\langle x^2\right\rangle^{1/2}\right).						
\end{eqnarray}
However, when $B(x)$ takes other forms,
e.g., $B(x)=\kappa x^n$ with $\kappa$
constant
and $n=3, 4, 5, \dots$, 
we cannot obtain the above relationship.

\subsection{Moment-based formulas for the Fokker-Planck coefficients}
\label{The Fokker-Planck equation with
the first- and second-order terms}

For the Fokker-Planck equation with
the first- and second-order terms
\begin{eqnarray}
\frac{\partial{F}}{\partial{t}}
=
-\frac{\partial{}}{\partial{x}}	
\left[A(x)F(x,t)\right]
+
\frac{\partial^2{}}{\partial{x^2}}	
\left[B(x)F(x,t)\right],
\label{Fokker-Planck equation with
	the first- and second-order terms}			 				
\end{eqnarray}	
we can readily obtain the equations for 
the first- and second-order
moments as
\begin{eqnarray}
&&\frac{\dee}{\dee t}\left\langle t\right\rangle
=
\left\langle A(x)\right\rangle,\\
&&\frac{\dee}{\dee t}\left\langle x^2\right\rangle
=
2\left\langle xA(x)\right\rangle
+
2\left\langle B(x)\right\rangle.		
\end{eqnarray}
If $A(x)=\kappa_x=\text{constant}$ 
and $B(x)=\kappa_{xx}=\text{constant}$,
the latter formulas become
\begin{eqnarray}
&&\kappa_x
=
\frac{\dee}{\dee t}
\left\langle x\right\rangle,\\
&&\kappa_{xx}
=
\frac{1}{2}
\frac{\dee}{\dee t}
\left[
\left\langle x^2\right\rangle
-
\left\langle x\right\rangle^2
\right]. 		
\end{eqnarray}
However, when $A(x)$ and $B(x)$ take other
forms, we cannot find the simple relationship 
between transport coefficients and statistical
quantities. For example,
if $A(x)=\kappa_x$ and $B(x)=\kappa_{xx}x$,
the coefficients take the form
\begin{eqnarray}
&&\kappa_x
=
\frac{\dee}{\dee t}
\left\langle x\right\rangle,\\
&&\kappa_{xx}
=
\frac{1}{2}
\frac{1}{\left\langle x\right\rangle}
\frac{\dee}{\dee t}
\left[
\left\langle x^2\right\rangle
-
\left\langle x\right\rangle^2
\right].	
\end{eqnarray}

\subsection{Moment-based coefficient formulas 
for some Fokker-Planck-type equations}
\label{Some other forms of the Fokker-Planck 
equation}

The Ornstein-Uhlenbeck processes are important in
physics, finance, data science, and other fields.
Its governing equation, which takes the following
form, is often used in related research
\begin{eqnarray}
\frac{\partial{F}}{\partial{t}}
=
\gamma\frac{\partial{}}{\partial{x}}	
\left(xF(x,t)\right)
+
Q
\frac{\partial^2{F}}{\partial{x^2}}	 	
\label{Ornstein-Uhlenbeck equation}			
\end{eqnarray}
Here, $\gamma$ and $Q$ are independent of the 
coordinate $x$.
Equation 
(\ref{Fokker-Planck equation with
	the first- and second-order terms})
reduces to Equation
(\ref{Ornstein-Uhlenbeck equation}) 
for  $A(x)=-\gamma x$ and 
$B(x)=Q$,
with $Q$ constant. 
Employing the same method,
we find the coefficient formula
for the Ornstein-Uhlenbeck equation
\begin{eqnarray}
&&\gamma=-\frac{\dee}{\dee t}
\ln
\left\langle x\right\rangle,
\label{gamma}\\
&&Q=\frac{1}{2}\frac{\dee}{\dee t}
\left\langle x^2\right\rangle
-\left\langle x^2\right\rangle
\frac{\dee}{\dee t}
\ln
\left\langle x\right\rangle
=\frac{1}{2}\frac{\dee}{\dee t}
\left\langle x^2\right\rangle
+\gamma\left\langle x^2\right\rangle,
\label{Q}
\end{eqnarray}
Equation (\ref{gamma})
indicates that the first-order coefficient 
$\gamma$ is a function of the 
expectation, and 
Equation (\ref{Q}) shows that
the second-order coefficient $Q$
is a function of the second-order moment
and $\gamma$. 
In addition, the Shimizu-Yamada model is 
a well-known formulation describing 
many-body systems with attractive coupling forces 
and time delays
\begin{eqnarray}
\frac{\partial{F}}{\partial{t}}
=
\frac{\partial{}}{\partial{x}}
\left\{
\left[
\gamma x
+\kappa(x-\left\langle x\right\rangle)
\right]F
\right\}	
+
Q
\frac{\partial^2{F}}{\partial{x^2}},		
\end{eqnarray}
where $\gamma, \kappa$ and $Q$ are independent 
of $x$.
Using the same method as in the previous subsection 
and starting from the  
latter model,
we derive
\begin{eqnarray}
&&\gamma=-\frac{\dee}{\dee t}
\ln
\left\langle x\right\rangle,
\\
&&Q=\frac{1}{2}\frac{\dee}{\dee t}
\left\langle x^2\right\rangle
+2\left\langle x^2\right\rangle\gamma
-
\left[
\left\langle x^2\right\rangle
-
\left\langle x\right\rangle^2
\right]
=
\frac{1}{2}\frac{\dee}{\dee t}
\left\langle x^2\right\rangle
+2\left\langle x^2\right\rangle\gamma
-
\sigma^2.
\end{eqnarray}
Here, $\sigma^2=\left\langle x^2\right\rangle
-
\left\langle x\right\rangle^2$. 
The first formula indicates that
the coefficient $\gamma$ takes the logarithmic
form of the first-order moment; 
the second formula shows that 
the second coefficient $Q$ 
is a function of the second-order moment
and the variance. 

The Desai-Zwanzig model is a well-known
description for
the order-disorder phase transition and takes the form
\begin{eqnarray}
\frac{\partial{F}}{\partial{t}}
=
-\frac{\partial{}}{\partial{x}}
\left[
ax-bx^3-
\kappa(x-\left\langle x\right\rangle)
\right]F	
+
Q
\frac{\partial^2{F}}{\partial{x^2}}. 			
\end{eqnarray}
Here, $a$, $b$, $\kappa$, and $Q$ are 
coefficients.
From the latter equation, we can obtain
\begin{eqnarray}
&&\frac{\dee}{\dee t}
\left\langle x\right\rangle
=a\left\langle x\right\rangle
-b\left\langle x^3\right\rangle,\\
&&\frac{\dee}{\dee t}
\left\langle x^2\right\rangle
=2\left[
a
\left\langle x^2\right\rangle
-b\left\langle x^4\right\rangle
-\kappa 
\left\langle x^2\right\rangle
+\kappa 
\left\langle x\right\rangle^2
\right]+2Q,\\
&&\frac{\dee}{\dee t}
\left\langle x^3\right\rangle
=3\left[
a
\left\langle x^3\right\rangle
-b\left\langle x^5\right\rangle
-\kappa 
\left\langle x^3\right\rangle
+\kappa
\left\langle x\right\rangle 
\left\langle x^2\right\rangle
\right]
+6Q\left\langle x\right\rangle,\\
&&\frac{\dee}{\dee t}
\left\langle x^4\right\rangle
=4\left[
a
\left\langle x^4\right\rangle
-b\left\langle x^6\right\rangle
-\kappa 
\left\langle x^4\right\rangle
+\kappa
\left\langle x\right\rangle 
\left\langle x^3\right\rangle
\right]
+12Q\left\langle x^2\right\rangle.
\end{eqnarray}
With Cramer's rule,
the formulas for $a$, $b$, $\kappa$, and $Q$
can be derived 
from the linear equations. 
These coefficients
are all functions of the moments.

The Klein-Kramers equation is 
another well-known model
in statistical mechanics and it takes the form 
\begin{eqnarray}
\frac{\partial{f}}{\partial{t}}
=
-v\frac{\partial{f}}{\partial{x}}
-\frac{V'(x)}{m}
\frac{\partial{f}}{\partial{v}}
+
\frac{\gamma}{m}
\frac{\partial{}}{\partial{v}}
\left(vf\right)
+\frac{\gamma k_B T}{m^2}
\frac{\partial^2{f}}{\partial{v^2}}.			
\end{eqnarray}
Here, $x$ and $v$ are variables, and $m$,
$\gamma$, $k_B$, and $T$ are constants. 
Using the definition of the moment 
\begin{eqnarray}
&&\left\langle x^iv^j \right\rangle
=\iint x^i v^jf(x, v, t)\dee x\dee v, 
\end{eqnarray}	
we obtain the following formulas
from the Klein-Kramers equation, 
\begin{eqnarray}
&&
\frac{\dee}{\dee t}
\left\langle x\right\rangle
=\left\langle v\right\rangle,\\
&&
\frac{\dee}{\dee t}
\left\langle v\right\rangle
=
\frac{1}{m}
\left\langle V'(x)\right\rangle
-
\frac{\gamma}{m}
\left\langle v\right\rangle.	
\end{eqnarray}
From the latter equations, we obtain
\begin{eqnarray}
\left\langle V'(x)\right\rangle
=m\frac{\dee}{\dee t}
\left\langle v\right\rangle
+\gamma
\frac{\dee}{\dee t}
\left\langle x\right\rangle
=\frac{\dee}{\dee t}
\left[
m\left\langle v\right\rangle
+\gamma
\left\langle x\right\rangle
\right].
\label{<V>}
\end{eqnarray}
If we set 
\begin{eqnarray}
Q_8=m\left\langle v\right\rangle
+\gamma
\left\langle x\right\rangle,
\end{eqnarray}
the coefficient can be rewritten as
\begin{eqnarray}
\left\langle V'(x)\right\rangle
=\frac{\dee}{\dee t}Q_8.
\end{eqnarray}
In addition, if $V'(x)$ is known,
we find
\begin{eqnarray}
\gamma
=
\frac{
\left\langle V'(x)\right\rangle
-m\frac{\dee}{\dee t}
\left\langle v\right\rangle
}{\frac{\dee}{\dee t}
\left\langle x\right\rangle}.
\end{eqnarray}

\section{Moment-based coefficient formula for
fractional transport equation}
\label{Moment-based Fractional transport equation 
coefficient}

The fractional transport equations, which 
describe
the anomalous diffusion of energetic charged particles
in turbulent magnetic fields, have been 
extensively explored over the past decades
\citep{MazurEA2000, RuffoloEA2003, NegreteEA2004b,
TrenchiEA2013, OliveiraEA2014, PerriEA2015,
ZimbardoEA2021}. 
Starting from the 
Continuous Time Random Walk method
\citep{MontrollWeis1965}, 
\citet{NegreteEA2004b} derived the well-known 
fractional equation
\begin{eqnarray}
\frac{\partial^{\alpha}{F}}{\partial{t^{\alpha}}}=
\chi\frac{\partial^{\beta}{F}}
{\partial{z^{\beta}}}.
\label{fractional transport equation with one 
order time derivative and
beta order spatial derivative-1tbetaz}	
\end{eqnarray}
Here, $0<\alpha<1$ and $0<\beta<2$, $F=F(z,t)$
is the distribution function of energetic 
particles, and $\chi$ is the transport coefficient.
In addition, the time derivative used 
in the latter equation is  
the Caputo derivative
\begin{eqnarray}
\frac{\partial^{\alpha}{F}}{\partial{t^{\alpha}}}
=\frac{1}{\Gamma(m-\alpha)}
\int_0^t d\tau
\left(t-\tau\right)^{m-\alpha-1}\frac{d^m}{d\tau^m}F,
\label{Dtxdelta}	
\end{eqnarray}
where $m$ is the ceiling of 
$\alpha$. 
The space derivative is defined according to
the Riesz definition 
\begin{eqnarray}
\frac{\partial^{\beta}{F}}		
{\partial{z^{\beta}}}=
-\frac{1}{2\cos\frac{\pi\beta}{2}}
\frac{1}{\Gamma(n-\beta)}
\frac{\dee^n}{\dee z^n}
\left(\int_{-\infty}^z
(z-\xi)^{n-\beta-1}F(\xi)\dee\xi
+(-1)^n\int_z^{+\infty}
(\xi-z)^{n-\beta-1}F(\xi)\dee\xi
\right)
\end{eqnarray}
Here, the left-sided and right-sided Weyl 
derivative are used.
In this section, we investigate 
the transport coefficient $\chi$ and 
derive its formula expressed in terms of 
the moments
of the distribution function
$F(z,t)$. 
Applying the Caputo derivative 
operator 
to
the $\alpha$-order moment 
of the distribution function
$F(z,t)$ yields
\begin{eqnarray}
\frac{\partial^{\alpha}{}}{\partial{t^{\alpha}}}
\left\langle |z|^{\alpha} \right\rangle
=\chi
\int_{-\infty}^{+\infty} dz 
|z|^{\alpha}\frac{1}{\Gamma(m-\alpha)}
\int_0^t d\tau
\left(t-\tau\right)^{m-\alpha-1}
\frac{d^m}{d\tau^m}F(z, \tau).
\label{Dtxdelta-1}			
\end{eqnarray}
where $m$ is the ceiling of $\alpha$. 
Similarly, we have
\begin{eqnarray}
\frac{\partial^{\beta}{}}		
{\partial{z^{\beta}}}
\left\langle |z|^{\alpha} \right\rangle
=&&
-
\int_{-\infty}^{+\infty}\dee z
|z|^{\alpha}
\frac{1}{2\cos\frac{\pi\beta}{2}}
\frac{1}{\Gamma(n-\beta)}
\frac{\dee^n}{\dee z^n}
\Bigg(\int_{-\infty}^z
(z-\xi)^{n-\beta-1}F(\xi)\dee\xi
\nonumber\\
&&
+(-1)^n\int_z^{+\infty}
(\xi-z)^{n-\beta-1}F(\xi)\dee\xi
\Bigg).
\label{moment for Riesz}
\end{eqnarray}
In the following subsection, we derive 
the formula for the coefficient $\chi$. 

\subsection{Moment-based coefficient formula for 
the fractional transport equation 
$\partial{F}/\partial{\MakeLowercase{t}}
=\chi
\partial^{\beta}{F}/
\partial{\MakeLowercase{x}^{\beta}}
$}
\label{Deducing the fractional transport equation coefficients}

The fractional transport equation
\begin{eqnarray}
\frac{\partial
{F}}{\partial{t}}=\chi \frac{\partial^{\beta}{F}}		
{\partial{z^{\beta}}}
\label{space negrete equation}			
\end{eqnarray} 
describes the evolution of 
the distribution function of energetic 
particles propagating in magnetic 
turbulence. The distribution function
of the fractional transport equation follows 
the L\'evy distribution.  
In this subsection, we derive the formula for
the coefficient $\chi$ in Equation
(\ref{space negrete equation}).

Combining Equations 
(\ref{moment for Riesz})
and 
(\ref{space negrete equation}), 
we find
\begin{eqnarray}
\int_{-\infty}^{+\infty}\dee z |z|^{\alpha}
\frac{\partial{F}}{\partial{t}}
&&=\frac{\dee}{\dee{t}}
\left\langle |z|^{\alpha} \right\rangle
=\chi\int_{-\infty}^{+\infty} dz |z|^{\alpha}
\frac{\partial^{\beta}{F}}
{\partial{z^{\beta}}}
\nonumber\\
&&
=\chi 
\int_{-\infty}^{+\infty} dz
|z|^{\alpha}
\left[-\frac{1}{2\cos\frac{\pi\beta}{2}}
\frac{1}{\Gamma(n-\beta)}
\frac{\dee^n}{\dee z^n}
\left(\int_{-\infty}^z
(z-\xi)^{n-\beta-1}F(\xi)\dee\xi
+(-1)^n\int_z^{+\infty}
(\xi-z)^{n-\beta-1}F(\xi)\dee\xi
\right)
\right]
\nonumber\\
&&
=
(-1)^{n+1}\frac{\chi}{2\cos \frac{\pi\beta}{2}}
\frac{\Gamma(\alpha+1)}{\Gamma(n-\beta)
\Gamma(\alpha-n+1)} 
\left[
I_1+(-1)^nI_2
\right]
\label{D_t^gamma z^alpha=C(I1+I2)}	
\end{eqnarray}
with
\begin{eqnarray}
I_1
&&=
\int_{-\infty}^{+\infty} dz
|z|^{\alpha-n}
\int_{-\infty}^z d\xi\left(z-\xi\right)
^{n-\beta-1}F(\xi),\\
I_2&&=
\int_{-\infty}^{+\infty} dz
|z|^{\alpha-n}
\int_z^{+\infty} 
d\xi\left(\xi-z\right)
^{n-\beta-1}F(\xi).		
\end{eqnarray}
The derivation  of 
the formulas $I_1$ and $I_2$ can be found in 
Appendix 
\ref{The derivation  of moment
	formula for the fractional transport 
	equation with the formulas $I_1$ and $I_2$},
and with the results obtained there,
the transport coefficient $\chi$
can be rewritten as
\begin{eqnarray}
\chi	
=2\cos \frac{\pi\beta}{2}
\frac{\Gamma(\alpha-n+1)}{\Gamma(\alpha+1)}
\frac{1}
{\Bigg(
	\frac{\Gamma(\beta-\alpha)}
	{\Gamma(n-\alpha)}
	\cos^2\frac{\alpha(\pi+2k\pi)}{2}
	-
	\frac{\Gamma(\alpha-n+1)}
	{\Gamma(\alpha-\beta+1)}
	\cos^2\frac{\beta(\pi+2k\pi)}{2}
	\Bigg)
	\left\langle 
	|z|^{\alpha-\beta}
	\right\rangle}
\frac{\dee}{\dee{t}}
\left\langle |z|^{\alpha} \right\rangle.
\end{eqnarray}
The latter formula shows that 
the coefficient $\chi$ is a function of 
$\left\langle |z|^{\alpha} \right\rangle$ and 
$\left\langle 
|z|^{\alpha-\beta}
\right\rangle$.

\subsection{Moment-based coefficient formula for fraction equation $\partial^{\alpha}{F}/\partial{t^{\alpha}}
=\chi\partial^2{F}/\partial{z^2}$}
\label{The transport coefficient of Equation alphat1z}

For $\beta=2$, Equation 
(\ref{fractional transport equation with one 
	order time derivative and
	beta order spatial derivative-1tbetaz}) 
becomes
\begin{eqnarray}
\frac{\partial^{\alpha}{F}}{\partial{t^{\alpha}}}=
\chi\frac{\partial^2{F}}
{\partial{z^2}}.
\end{eqnarray}
In this subsection, we derive 
the transport coefficient 
$\chi$ in the latter equation. 
Using the same procedure as in the 
preceding subsections yields
the following formula
\begin{eqnarray}
	\frac{d^{\alpha}}{dt^{\alpha}}\langle  z^{2}\rangle
	=2\kappa_{\alpha}
	\int_{-\infty}^{\infty}dz F(z).
	\label{d alpha z2 dt alpha}
\end{eqnarray} 
Given the normalization condition 
and assuming the distribution function 
is even, we have
\begin{eqnarray}
	\int_{-\infty}^{+\infty}dz F(z ,t)
	=2\int_{0}^{+\infty}dz F(z ,t)=1, 
	\label{normaliztion condition}
\end{eqnarray}
From Equations 
(\ref{d alpha z2 dt alpha})
and 
(\ref{normaliztion condition}),
we find 
\begin{eqnarray}
\chi=\frac{1}{2}\frac{d^{\alpha}}
{dt^{\alpha}}\langle  z^{2}\rangle,
\end{eqnarray}  
which indicates that 
the transport coefficient $\chi$ is a function
of the second-order moment.
As done in Section
(\ref{Deducing the fractional transport equation coefficients}),
the power-law dependence of the second-order 
moment of $z$
can be obtained 
\begin{eqnarray}
	\langle  z^{2}\rangle\sim t^{\alpha}. 
	\label{power law for alphat2z}
\end{eqnarray}
If $\alpha=1$, the latter formula describes 
a normal Markovian
process; for $1<\alpha<2$ and $0<\alpha<1$, 
it corresponds to 
the subdiffusion and superdiffusion, 
respectively. 

\subsection{Moment-based coefficient formula for 
fractional equation $\partial{F}/\partial{t}
=\chi_{\beta}
\partial^{\beta}{F}/\partial{z^{\beta}}
+
\chi_{\alpha}
\partial^{\alpha}{F}/\partial{z^{\alpha}}
$}
\label{The transport coefficient of Equation 1}

In this subsection, we investigate 
the transport coefficients of 
the following fractional equation
\begin{eqnarray}
\frac{\partial{F}}{\partial{t}}=
\chi_{\beta}\frac{\partial^{\beta}{F}}
{\partial{z^{\beta}}}
+
\chi_{\alpha}\frac{\partial^{\alpha}{F}}
{\partial{z^{\alpha}}},
\label{fractional transport equation with 
two terms}		
\end{eqnarray}
where $0<\alpha, \beta<2$.
Applying the Fourier transform
\begin{eqnarray}
\hat{F}(k)=\int_{-\infty}^{+\infty}F(z)e^{-ikz}
\dee z,	
\end{eqnarray}
we find that 
Equation
(\ref{fractional transport equation with 
two terms})
becomes
\begin{eqnarray}
\frac{\partial{\hat{F}}}{\partial{t}}
=
-\kappa_{\beta}|k|^{\beta}\hat{F}
-\kappa_{\alpha}|k|^{\alpha}\hat{F}.	
\end{eqnarray}
By defining the following formulas for the moments 
in Fourier space
\begin{eqnarray}
&&
\left\langle 
E
\right\rangle
=
\int_{-\infty}^{+\infty}\dee k \hat{F}(k),\\
&&
\left\langle 
|k|^{\alpha}
\right\rangle
=
\int_{-\infty}^{+\infty}\dee k \hat{F}(k)|k|^{\alpha},	
\end{eqnarray}
we obtain the zeroth- and $\beta$-th order
moment equation as
\begin{eqnarray}
&&
\frac{\dee}{\dee t}
\left\langle 
E
\right\rangle
=
-\kappa_{\beta}
\left\langle 
|k|^{\beta}
\right\rangle
-\kappa_{\alpha}
\left\langle 
|k|^{\alpha}
\right\rangle,\\
&&
\frac{\dee}{\dee t}
\left\langle 
|k|^{\beta}
\right\rangle
=
-\kappa_{\beta}
\left\langle 
|k|^{2\beta}
\right\rangle
-\kappa_{\alpha}
\left\langle 
|k|^{\beta+\alpha}
\right\rangle.		
\end{eqnarray}
Solving the two linear equations
gives
\begin{eqnarray}
&&
\kappa_{\beta}
=
\frac{
	\left\langle |k|^{\alpha}\right\rangle
	\frac{\dee}{\dee t}
	\left\langle |k|^{\beta}\right\rangle
	-
	\left\langle |k|^{\beta+\alpha}\right\rangle
	\frac{\dee}{\dee t}
	\left\langle 
	E
	\right\rangle
}
{
	\left\langle |k|^{\beta}\right\rangle
	\left\langle |k|^{\beta+\alpha}\right\rangle
	-
	\left\langle |k|^{\alpha}\right\rangle
	\left\langle |k|^{2\beta}\right\rangle
},
\label{kalpha for two terms}\\
&&
\kappa_{\alpha}
=
\frac{
	\left\langle |k|^{2\beta}\right\rangle
	\frac{\dee}{\dee t}
	\left\langle 
	E
	\right\rangle
	-
	\left\langle |k|^{\beta}\right\rangle
	\frac{\dee}{\dee t}
	\left\langle |k|^{\beta}\right\rangle	
}
{
	\left\langle |k|^{\beta}\right\rangle
	\left\langle |k|^{\beta+\alpha}\right\rangle
	-
	\left\langle |k|^{\alpha}\right\rangle
	\left\langle |k|^{2\beta}\right\rangle	
}.
\label{kbeta for two terms}	
\end{eqnarray}
Since the moments  
$\left\langle |k|^{\alpha}\right\rangle$, 
$\left\langle |k|^{\beta}\right\rangle$,
and other moments 
are all functions of time, 
the coefficients are also 
functions of time. 
Similarly, 
after applying the Fourier transform
to the fractional equation, 
\begin{eqnarray}
\frac{\partial
{F}}{\partial{t}}=\chi \frac{\partial^{\alpha}{F}}			
{\partial{z^{\alpha}}},
\end{eqnarray} 
we derive
\begin{eqnarray}
\frac{\partial{\hat{F}}}{\partial{t}}
=
-\chi|k|^{\alpha}\hat{F}. 	
\end{eqnarray}
The corresponding zeroth moment equation can be 
found
\begin{eqnarray}
\frac{\dee}{\dee t}
\left\langle 
E
\right\rangle
=
-\chi
\left\langle 
|k|^{\alpha}
\right\rangle.	
\end{eqnarray}
Thus, the formula for the transport coefficient
can be obtained 
\begin{eqnarray}
\kappa
=
-\frac{1}{\left\langle 
	|k|^{\alpha}
	\right\rangle}
\frac{\dee}{\dee t}
\left\langle 
E
\right\rangle. 
\label{kalpha for one terms}
\end{eqnarray}
Because $\left\langle 
|k|^{\alpha}
\right\rangle$ is a function of time,
the coefficient $\kappa$ is dependent on time.

\section{SUMMARY AND CONCLUSION}
\label{SUMMARY AND CONCLUSION}

Transport coefficients are important 
physical quantities describing stochastic
processes. 
In previous studies, the spatial transport 
coefficients (including parallel and 
perpendicular) have been assumed constant
and can be expressed in terms of 
various statistical quantities, such as the mean and
variance\citep{wq2023}.
However, the momentum transport coefficients
have not been explored. In this paper, we 
investigate whether the momentum transport 
formulas can be expresses in terms of 
moments.  
In addition,  
pitch-angle scattering coefficient
depend on the variables. 
Whether a relationship exists between
variable transport coefficients 
and statistical quantities remain unclear.  
In this paper, we explore this issue.

Employing the method from our previous 
studies,  
we derive 
the coefficients
for momentum, pitch-angle,
and spatial transport 
in terms of the moments. 
Starting with the momentum transport equations
with variable coefficients, i.e., 
Equation (\ref{Fokker-Planck equation used}),
we show that the coefficients 
are the logarithmic functions of
statistical quantities.  
For the momentum equation containing only one term,
e.g., the convection term or 
the diffusion term,
the transport coefficients 
$\mathcal{K}_p$, $\mathcal{K}_1$ and 
$\mathcal{K}_2$ are all logarithmic 
functions of the mathematical expectation.
However, if the momentum transport equations
contain two terms, 
either the first and second diffusion terms,
or the convection term and a diffusion term,
the transport coefficients are all 
logarithmic functions of some new statistical 
quantities. 
The results show that 
the momentum transport coefficients 
take logarithmic forms, which are different
from those of the 
constant spatial transport coefficients.
Similarly, we find that 
the pitch-angle diffusion coefficient
also takes the logarithmic form,
which is identical to the result obtained
by \citet{Tautz2013}. 
In addition, 
for the first-order
Fokker-Planck equation with the coefficient
$A(x)=\kappa_x x$ and
the second-order Fokker-Planck equation
with the coefficient
$B(x)=\kappa_{xx} x^2$, 
the coefficients $\kappa_x$ and
$\kappa_{xx}$ also take a logarithmic
form. 

There are many Fokker-Planck-type 
transport models in space plasma physics,
statistical physics, finance,
and other fields.  
The Ornstein-Uhlenbeck equation,
where the first coefficient is not constant
and 
the second coefficient is independent of $x$,
is well-known 
in finance and data science. 
Using the same procedure, we find that
the first coefficient $\gamma$ 
is a logarithmic function
of the expectation, 
and the second coefficient $Q$ 
is a function of the second-order moment
and the variance. 
The Shimizu-Yamada equation,
the Desai-Zwanzig equation
and the Klein-Kramers equation
are all well-known in statistical 
mechanics, and 
their coefficients are all functions
of the moments. 

Anomalous diffusion processes, 
including 
superdiffusion and subdiffusion,
have been extensively investigated 
in recent decades.
The fractional transport equation
with the Caputo time derivative
and the Riesz space derivative,
derived by \citet{NegreteEA2004b}, 
is one of the best-known models. 
In this paper, we find that 
the transport coefficients
of the fractional equation can be
expressed in terms of the moments.

The coefficient formulas derived in this paper
are the simple theoretical results.
Because we assume an isotropic form  $D_{\mu\mu}=D(1-\mu^2)$ with a constant $D$,
the results obtained in this paper are 
only applicable to isotropic pitch-angle  
scattering and to pitch-angle diffusion that is
almost independent of particle speed. 
In addition, because 
the mixed-derivative terms with
$D_{\mu p}$ or $D_{p\mu}$
are neglected,
the formulas obtained in this paper 
are only applicable to cases in which 
the coupling effects between pitch-angle
and particle momentum are very weak
relative to the other physical effects.  
Additionally, for simplicity,
we assume that $D_{pp}=N p^2D_{\mu\mu}$ with
$N$ constant. Actually, the parameter $N$ is a
function of particle speed and other variables.
Thus, the results obtained in this paper 
are only applicable to the cases in which $D_{pp}$
is approximately quadratic in particle 
momentum. 
This paper is only the starting 
point for our research on this topic. 
As the next step, 
we will explore cases that include 
the terms with $D_{\mu p}$ or $D_{p\mu}$, 
and cases in which 
$D_{\mu\mu}$ depends on particle momentum. 
In addition, 
we will explore the formulas for the
momentum transport coefficients 
when $D_{pp}$, $D_{p}$ and 
$D_{\mu\mu}$ depend on other variables
or take other forms.

\begin{acknowledgments}
	\nnsfc{\WangjfNNSFC, \QinNNSFCouter}
	\szstp{\QinSZdeeplearning, \QinSZdisaster}
	\nkrdpc{\GuoNKRDPC, \ShenNKRDPC}
	\szkllp{\FengLAB}
	\referee
\end{acknowledgments}

\renewcommand{\theequation}{\Alph{section}-\arabic{equation}}
\setcounter{equation}{0}  
\begin{appendices}
	
\section{The derivation of the momentum transport equation}
\label{The derivation of the momentum transport equation}	

The Fokker–Planck equation with momentum 
transport terms is is given by 
\citep{Schlickeiser2002}
\begin{eqnarray}
\frac{\partial{f}}{\partial{t}}
&=&
-v\mu\frac{\partial{f}}{\partial{z}}
+\frac{\partial{}}{\partial{\mu}}
\left(D_{\mu\mu}
\frac{\partial{f}}{\partial{\mu}}\right)
-\frac{1}{p^2}\frac{\partial{}}{\partial{p}}
(p^2 D_{p}f)
+\frac{1}{p^2}\frac{\partial{}}{\partial{p}}
\left(p^2 D_{pp}\frac{\partial{f}}{\partial{p}}
\right)
\label{momentum Fokker-Planck equation}		
\end{eqnarray}
with
\begin{eqnarray}
&&D_{p}=pA(\mu),\\
&&D_{pp}=
N p^2D_{\mu\mu}. 		
\end{eqnarray}
Here, $N$ is constant. 
In fact, the coefficients
can take more general forms, e.g., 
\begin{eqnarray}
&&D_{p}=p^nA(\mu),\\
&&D_{pp}=
N p^mD_{\mu\mu},			
\end{eqnarray} 
and these will be explored in future work. 

Decomposing the distribution function 
$f=f(z, p, \mu, t)$
into the isotropic part $F=F(z, p, t)$
and the anisotropic part
$g=g(z, p, \mu, t)$ yields
\begin{eqnarray}
f=F+g.
\label{f=F+g}
\end{eqnarray}
Here, the following formulas hold
\begin{eqnarray}
&&\int_{-1}^{+1}g\dee\mu=0,\\
&& \int_{-1}^{+1}f\dee\mu=F.
\end{eqnarray}
Inserting Equation (\ref{f=F+g}) into 
Equation
(\ref{momentum Fokker-Planck equation})
gives
\begin{eqnarray}
\frac{\partial{F}}{\partial{t}}
+
\frac{\partial{g}}{\partial{t}}
&=&
-v\mu\frac{\partial{F}}{\partial{z}}
-v\mu\frac{\partial{g}}{\partial{z}}
+\frac{\partial{}}{\partial{\mu}}
\left(D_{\mu\mu}
\frac{\partial{g}}{\partial{\mu}}\right)
-\frac{1}{p^2}\frac{\partial{}}{\partial{p}}
\left(p^{3}A(\mu)F\right)
-\frac{1}{p^2}\frac{\partial{}}{\partial{p}}
\left(p^{3}A(\mu)g\right)
\nonumber\\
&&
+N\frac{1}{p^2}\frac{\partial{}}{\partial{p}}
\left(
p^{4}D_{\mu\mu}	
\frac{\partial{F}}{\partial{p}}	
\right)
+N\frac{1}{p^2}\frac{\partial{}}{\partial{p}}
\left(p^{4}D_{\mu\mu}	
\frac{\partial{g}}{\partial{p}}	
\right)	
\label{momentum equation with F+g}	
\end{eqnarray}
By integrating the latter equation 
over $\mu$ from $-1$ to $+1$, we obtain
the following momentum equation
\begin{align}
\frac{\partial{F}}{\partial{t}}
=-\kappa_p
\frac{1}{p^2}\frac{\partial{}}{\partial{p}}
\left(p^{3}F\right)
+
\kappa_{pp}^{(0)}
\frac{1}{p^2}\frac{\partial{}}{\partial{p}}
\left(p^{4}
\frac{\partial{F}}{\partial{p}}\right)	
+T_1
+T_2
+T_3,	
\label{momentum equation with 3T}		
\end{align}
where the constant coefficients are defined as 
\begin{align}
&\kappa_p=	\frac{1}{2}\int_{-1}^{+1}\dee\mu A(\mu),\\
&\kappa_{pp}^{(0)}=\frac{N}{2}\int_{-1}^{+1}\dee\mu
D_{\mu\mu}. 		
\end{align}
The three terms \(T_1\), \(T_2\), and \(T_3\)
are given by
\begin{align}
&T_1=-\frac{v}{2}\frac{\partial{}}{\partial{z}}
\int_{-1}^{+1}\dee\mu\mu g,\\
&T_2=-\frac{1}{2}
\frac{1}{p^2}\frac{\partial{}}{\partial{p}}
p^{3}\int_{-1}^{+1}\dee\mu
A(\mu)g,\\
&T_3=\frac{1}{p^2}
\frac{\partial{}}{\partial{p}}
p^{4}\frac{\partial{}}{\partial{p}}	
\frac{N}{2}\int_{-1}^{+1}\dee\mu
D_{\mu\mu}	g.			
\end{align}
Integrating Equation
(\ref{momentum equation with F+g})
over $\mu$ from $-1$ to $\mu$ gives
\begin{eqnarray}
\frac{\partial{g}}{\partial{\mu}}
=\frac{1}{D_{\mu\mu}}
\Phi(\mu)	
\label{g/mu with Phi}
\end{eqnarray}
with the function
\begin{eqnarray}
\Phi(\mu)=&&
\Bigg[\frac{\partial{F}}{\partial{t}}(\mu+1)
+
\frac{\partial{}}{\partial{t}}
\int_{-1}^{\mu}\dee \mu g
\nonumber\\
&&
+v\frac{\mu^2-1}{2}\frac{\partial{F}}{\partial{z}}
+v\frac{\partial{}}{\partial{z}}
\int_{-1}^{\mu}\dee \mu \mu g
+\frac{1}{p^2}\frac{\partial{}}{\partial{p}}
p^{3} F
\int_{-1}^{\mu}\dee \mu A(\mu)
+\frac{1}{p^2}\frac{\partial{}}{\partial{p}}
p^{3}
\int_{-1}^{\mu}\dee \mu
A(\mu)g
\nonumber\\
&&
-\frac{1}{p^2}\frac{\partial{}}{\partial{p}}
\left(
p^{4}\frac{\partial{F}}{\partial{p}}	
N\int_{-1}^{\mu}\dee \mu
D_{\mu\mu}
\right)	
-N\frac{1}{p^2}\frac{\partial{}}{\partial{p}}
\left(
p^{4}\frac{\partial{}}{\partial{p}}
\int_{-1}^{\mu}\dee \mu
D_{\mu\mu}	g
\right)
\Bigg]. 			
\end{eqnarray}
From formula
(\ref{g/mu with Phi}) we can find 
\begin{eqnarray}
g(\mu)=\Bigg(\int_{-1}^{\mu}\dee\mu
\frac{1}{D_{\mu\mu}}
-\frac{1}{2}\int_{-1}^{+1}\dee\mu
\int_{-1}^{\mu}\dee\mu
\frac{1}{D_{\mu\mu}}\Bigg)
\Phi(\mu).		
\end{eqnarray}

\subsection{Deriving the formula for $T_1$}
\label{Deriving the formula of T1}

The term \(T_1\) can be rewritten as
\begin{align}
T_1=-\frac{v}{2}\frac{\partial{}}{\partial{z}}
\int_{-1}^{+1}\dee\mu\mu g
=\sum_{n=1}^8Z_n			
\end{align}
with
\begin{align}
&Z_1=-\frac{v}{2}\frac{\partial{}}{\partial{z}}
\int_{-1}^{+1}\dee\mu\mu 
\int_{-1}^{\mu}\dee\mu
\frac{1}{D_{\mu\mu}}
\frac{\partial{F}}{\partial{t}}(\mu+1)
\nonumber\\
&
Z_2=-\frac{v}{2}\frac{\partial{}}{\partial{z}}
\int_{-1}^{+1}\dee\mu\mu 
\int_{-1}^{\mu}\dee\mu
\frac{1}{D_{\mu\mu}}
\frac{\partial{}}{\partial{t}}
\int_{-1}^{\mu}\dee \mu g
\nonumber\\
&
Z_3=-\frac{v}{2}\frac{\partial{}}{\partial{z}}
\int_{-1}^{+1}\dee\mu\mu 
\int_{-1}^{\mu}\dee\mu
\frac{1}{D_{\mu\mu}}
v\frac{\mu^2-1}{2}\frac{\partial{F}}{\partial{z}}
\nonumber\\
&
Z_4=-\frac{v}{2}\frac{\partial{}}{\partial{z}}
\int_{-1}^{+1}\dee\mu\mu 
\int_{-1}^{\mu}\dee\mu
\frac{1}{D_{\mu\mu}}
v\frac{\partial{}}{\partial{z}}
\int_{-1}^{\mu}\dee \mu \mu g
\nonumber\\
&
Z_5=-\frac{v}{2}\frac{\partial{}}{\partial{z}}
\int_{-1}^{+1}\dee\mu\mu 
\int_{-1}^{\mu}\dee\mu
\frac{1}{D_{\mu\mu}}
\frac{1}{p^2}\frac{\partial{}}{\partial{p}}
p^{3} F
\int_{-1}^{\mu}\dee \mu A(\mu)
\nonumber\\
&
Z_6=-\frac{v}{2}\frac{\partial{}}{\partial{z}}
\int_{-1}^{+1}\dee\mu\mu 
\int_{-1}^{\mu}\dee\mu
\frac{1}{D_{\mu\mu}}
\frac{1}{p^2}\frac{\partial{}}{\partial{p}}
p^{3}
\int_{-1}^{\mu}\dee \mu
A(\mu)g
\nonumber\\
&
Z_7=\frac{v}{2}\frac{\partial{}}{\partial{z}}
\int_{-1}^{+1}\dee\mu\mu 
\int_{-1}^{\mu}\dee\mu
\frac{1}{D_{\mu\mu}}
\frac{1}{p^2}\frac{\partial{}}{\partial{p}}
p^{4}\frac{\partial{F}}{\partial{p}}	
N\int_{-1}^{\mu}\dee \mu
D_{\mu\mu}	
\nonumber\\
&
Z_8=\frac{v}{2}\frac{\partial{}}{\partial{z}}
\int_{-1}^{+1}\dee\mu\mu 
\int_{-1}^{\mu}\dee\mu
\frac{1}{D_{\mu\mu}}
N\frac{1}{p^2}\frac{\partial{}}{\partial{p}}
p^{4}\frac{\partial{}}{\partial{p}}
\int_{-1}^{\mu}\dee \mu
D_{\mu\mu}	g			
\end{align}
In this section, we retain only the terms 
involving 
momentum derivative operator
$\partial{}/\partial{p}$,
and neglect the vanishing terms.
In addition, we set the third-order and higher 
derivative terms to zero, retaining only the 
lower-order terms.
Based on these requirements, we obtain 
\begin{eqnarray}
&&Z_1=-\frac{v}{2}\frac{\partial{}}{\partial{z}}
\int_{-1}^{+1}\dee\mu\mu 
\int_{-1}^{\mu}\dee\mu
\frac{1}{D_{\mu\mu}}
\frac{\partial{F}}{\partial{t}}(\mu+1)=0,
\label{Z1}\\		
&&Z_2=-\frac{v}{2}\frac{\partial{}}{\partial{z}}
\int_{-1}^{+1}\dee\mu\mu 
\int_{-1}^{\mu}\dee\mu
\frac{1}{D_{\mu\mu}}
\frac{\partial{}}{\partial{t}}
\int_{-1}^{\mu}\dee \mu g=0,\\	
&&Z_3=-\frac{v}{2}\frac{\partial{}}{\partial{z}}
\int_{-1}^{+1}\dee\mu\mu 
\int_{-1}^{\mu}\dee\mu
\frac{1}{D_{\mu\mu}}
v\frac{\mu^2-1}{2}\frac{\partial{F}}{\partial{z}}=0,\\
&&Z_4=-\frac{v}{2}\frac{\partial{}}{\partial{z}}
\int_{-1}^{+1}\dee\mu\mu 
\int_{-1}^{\mu}\dee\mu
\frac{1}{D_{\mu\mu}}
v\frac{\partial{}}{\partial{z}}
\int_{-1}^{\mu}\dee \mu \mu g=0,\\	
&&Z_5=-\frac{v}{2}\frac{\partial{}}{\partial{z}}
\int_{-1}^{+1}\dee\mu\mu 
\int_{-1}^{\mu}\dee\mu
\frac{1}{D_{\mu\mu}}
\frac{1}{p^2}\frac{\partial{}}{\partial{p}}
p^{3} F
\int_{-1}^{\mu}\dee \mu A(\mu)
=-\kappa_{zp}^{(1)}	
\frac{1}{p}\frac{\partial{}}{\partial{p}}
p^{3}\frac{\partial{F}}{\partial{z}},\\ 	
&&Z_6=-\frac{v}{2}\frac{\partial{}}{\partial{z}}
\int_{-1}^{+1}\dee\mu\mu 
\int_{-1}^{\mu}\dee\mu
\frac{1}{D_{\mu\mu}}
\frac{1}{p^2}\frac{\partial{}}{\partial{p}}
p^{3}
\int_{-1}^{\mu}\dee \mu
A(\mu)g=0,\\		
&&Z_7=\frac{v}{2}\frac{\partial{}}{\partial{z}}
\int_{-1}^{+1}\dee\mu\mu 
\int_{-1}^{\mu}\dee\mu
\frac{1}{D_{\mu\mu}}
\frac{1}{p^2}\frac{\partial{}}{\partial{p}}
p^{4}\frac{\partial{F}}{\partial{p}}	
N\int_{-1}^{\mu}\dee \mu
D_{\mu\mu}
=0,\\	
&&Z_8=\frac{v}{2}\frac{\partial{}}{\partial{z}}
\int_{-1}^{+1}\dee\mu\mu 
\int_{-1}^{\mu}\dee\mu
\frac{1}{D_{\mu\mu}}
N\frac{1}{p^2}\frac{\partial{}}{\partial{p}}
p^{4}\frac{\partial{}}{\partial{p}}
\int_{-1}^{\mu}\dee \mu
D_{\mu\mu}	g=0	
\label{Z8}	
\end{eqnarray}
with
\begin{eqnarray}
\kappa_{zp}^{(1)}=\frac{1}{2}
\int_{-1}^{+1}\dee\mu\mu 
\int_{-1}^{\mu}\dee\mu
\frac{1}{D_{\mu\mu}}
\int_{-1}^{\mu}\dee \mu A(\mu). 
\label{kappa_zp1}				
\end{eqnarray}
Combining Equations 
(\ref{Z1})-(\ref{kappa_zp1}), we derive
\begin{eqnarray}
T_1=-\kappa_{zp}^{(1)}	
\frac{1}{p}\frac{\partial{}}{\partial{p}}
p^{3}\frac{\partial{F}}{\partial{z}}. 	 	
\end{eqnarray}

\subsection{$T_2$}

The term $T_2$ can be rewritten as
\begin{eqnarray}
T_2&=&-\frac{1}{2}
\frac{1}{p^2}\frac{\partial{}}{\partial{p}}
p^{3}\int_{-1}^{+1}\dee\mu
A(\mu)g	
=\sum_{n=1}^8X_n			
\end{eqnarray}
with
\begin{eqnarray}
&&X_1=-\frac{1}{2}
\frac{1}{p^2}\frac{\partial{}}{\partial{p}}
p^{3}
\int_{-1}^{+1}\dee\mu
A(\mu)\Bigg(\int_{-1}^{\mu}\dee\mu
\frac{1}{D_{\mu\mu}}
-\frac{1}{2}\int_{-1}^{+1}\dee\mu
\int_{-1}^{\mu}\dee\mu
\frac{1}{D_{\mu\mu}}\Bigg)
\frac{\partial{F}}{\partial{t}}(\mu+1),
\\
&&
X_2=-\frac{1}{2}
\frac{1}{p^2}\frac{\partial{}}{\partial{p}}
p^{3}
\int_{-1}^{+1}\dee\mu
A(\mu)\Bigg(\int_{-1}^{\mu}\dee\mu
\frac{1}{D_{\mu\mu}}
-\frac{1}{2}\int_{-1}^{+1}\dee\mu
\int_{-1}^{\mu}\dee\mu
\frac{1}{D_{\mu\mu}}\Bigg)
\frac{\partial{}}{\partial{t}}
\int_{-1}^{\mu}\dee \mu g,
\\
&&
X_3=-\frac{1}{2}
\frac{1}{p^2}\frac{\partial{}}{\partial{p}}
p^{3}
\int_{-1}^{+1}\dee\mu
A(\mu)\Bigg(\int_{-1}^{\mu}\dee\mu
\frac{1}{D_{\mu\mu}}
-\frac{1}{2}\int_{-1}^{+1}\dee\mu
\int_{-1}^{\mu}\dee\mu
\frac{1}{D_{\mu\mu}}\Bigg)
v\frac{\mu^2-1}{2}\frac{\partial{F}}{\partial{z}},
\\
&&
X_4=-\frac{1}{2}
\frac{1}{p^2}\frac{\partial{}}{\partial{p}}
p^{3}
\int_{-1}^{+1}\dee\mu
A(\mu)\Bigg(\int_{-1}^{\mu}\dee\mu
\frac{1}{D_{\mu\mu}}
-\frac{1}{2}\int_{-1}^{+1}\dee\mu
\int_{-1}^{\mu}\dee\mu
\frac{1}{D_{\mu\mu}}\Bigg)
v\frac{\partial{}}{\partial{z}}
\int_{-1}^{\mu}\dee \mu \mu g,
\\
&&
X_5=\frac{1}{2}
\frac{1}{p^2}\frac{\partial{}}{\partial{p}}
p^{3}
\int_{-1}^{+1}\dee\mu
A(\mu)\Bigg(\int_{-1}^{\mu}\dee\mu
\frac{1}{D_{\mu\mu}}
-\frac{1}{2}\int_{-1}^{+1}\dee\mu
\int_{-1}^{\mu}\dee\mu
\frac{1}{D_{\mu\mu}}\Bigg)
\frac{1}{p^2}\frac{\partial{}}{\partial{p}}
p^{3} F
\int_{-1}^{\mu}\dee \mu A(\mu),
\\
&&
X_6=\frac{1}{2}
\frac{1}{p^2}\frac{\partial{}}{\partial{p}}
p^{3}
\int_{-1}^{+1}\dee\mu
A(\mu)\Bigg(\int_{-1}^{\mu}\dee\mu
\frac{1}{D_{\mu\mu}}
-\frac{1}{2}\int_{-1}^{+1}\dee\mu
\int_{-1}^{\mu}\dee\mu
\frac{1}{D_{\mu\mu}}\Bigg)
\frac{1}{p^2}\frac{\partial{}}{\partial{p}}
p^{3}
\int_{-1}^{\mu}\dee \mu
A(\mu)g,
\\
&&
X_7=\frac{1}{2}
\frac{1}{p^2}\frac{\partial{}}{\partial{p}}
p^{3}
\int_{-1}^{+1}\dee\mu
A(\mu)\Bigg(\int_{-1}^{\mu}\dee\mu
\frac{1}{D_{\mu\mu}}
-\frac{1}{2}\int_{-1}^{+1}\dee\mu
\int_{-1}^{\mu}\dee\mu
\frac{1}{D_{\mu\mu}}\Bigg)
\frac{1}{p^2}\frac{\partial{}}{\partial{p}}
p^{4}\frac{\partial{F}}{\partial{p}}	
N\int_{-1}^{\mu}\dee \mu
D_{\mu\mu}, 	
\\
&&
X_8=\frac{1}{2}
\frac{1}{p^2}\frac{\partial{}}{\partial{p}}
p^{3}
\int_{-1}^{+1}\dee\mu
A(\mu)\Bigg(\int_{-1}^{\mu}\dee\mu
\frac{1}{D_{\mu\mu}}
-\frac{1}{2}\int_{-1}^{+1}\dee\mu
\int_{-1}^{\mu}\dee\mu
\frac{1}{D_{\mu\mu}}\Bigg)
N\frac{1}{p^2}\frac{\partial{}}{\partial{p}}
p^{4}\frac{\partial{}}{\partial{p}}
\int_{-1}^{\mu}\dee \mu
D_{\mu\mu}	g				
\end{eqnarray}
Using the same methods and 
requirements as in the latter subsection,
we obtain
\begin{eqnarray}
&&X_1=-\frac{1}{2}
\int_{-1}^{+1}\dee\mu
A(\mu)\Bigg(\int_{-1}^{\mu}\dee\mu
\frac{1}{D_{\mu\mu}}
-\frac{1}{2}\int_{-1}^{+1}\dee\mu
\int_{-1}^{\mu}\dee\mu
\frac{1}{D_{\mu\mu}}\Bigg)
\frac{\partial{F}}{\partial{t}}(\mu+1)
=-\kappa_{pt}^{(2)}
\frac{1}{p^2}\frac{\partial{}}{\partial{p}}
p^{3}	
\frac{\partial{F}}{\partial{t}},\\
&&X_2=-\frac{1}{2}
\frac{1}{p^2}\frac{\partial{}}{\partial{p}}
p^{3}
\int_{-1}^{+1}\dee\mu
A(\mu)\Bigg(\int_{-1}^{\mu}\dee\mu
\frac{1}{D_{\mu\mu}}
-\frac{1}{2}\int_{-1}^{+1}\dee\mu
\int_{-1}^{\mu}\dee\mu
\frac{1}{D_{\mu\mu}}\Bigg)
\frac{\partial{}}{\partial{t}}
\int_{-1}^{\mu}\dee \mu g=0,\\	
&&X_3=-\frac{1}{2}
\frac{1}{p^2}\frac{\partial{}}{\partial{p}}
p^{3}
\int_{-1}^{+1}\dee\mu
A(\mu)\Bigg(\int_{-1}^{\mu}\dee\mu
\frac{1}{D_{\mu\mu}}
-\frac{1}{2}\int_{-1}^{+1}\dee\mu
\int_{-1}^{\mu}\dee\mu
\frac{1}{D_{\mu\mu}}\Bigg)
v\frac{\mu^2-1}{2}\frac{\partial{F}}{\partial{z}}
\nonumber\\
&&
=-\kappa_{pz}^{(2)}
\frac{1}{p^2}\frac{\partial{}}{\partial{p}}
p^{4}
\frac{\partial{F}}{\partial{z}},\\		
&&X_4=-\frac{1}{2}
\frac{1}{p^2}\frac{\partial{}}{\partial{p}}
p^{3}
\int_{-1}^{+1}\dee\mu
A(\mu)\Bigg(\int_{-1}^{\mu}\dee\mu
\frac{1}{D_{\mu\mu}}
-\frac{1}{2}\int_{-1}^{+1}\dee\mu
\int_{-1}^{\mu}\dee\mu
\frac{1}{D_{\mu\mu}}\Bigg)
v\frac{\partial{}}{\partial{z}}
\int_{-1}^{\mu}\dee \mu \mu g=0,\\
&&X_5=\frac{1}{2}
\frac{1}{p^2}\frac{\partial{}}{\partial{p}}
p^{3}
\int_{-1}^{+1}\dee\mu
A(\mu)\Bigg(\int_{-1}^{\mu}\dee\mu
\frac{1}{D_{\mu\mu}}
-\frac{1}{2}\int_{-1}^{+1}\dee\mu
\int_{-1}^{\mu}\dee\mu
\frac{1}{D_{\mu\mu}}\Bigg)
\frac{1}{p^2}\frac{\partial{}}{\partial{p}}
p^{3} F
\int_{-1}^{\mu}\dee \mu A(\mu)	
\nonumber\\
&&=\kappa_{pp}^{(2)}
\frac{1}{p^2}\frac{\partial{}}{\partial{p}}
p
\frac{\partial{}}{\partial{p}}
p^{3} F,\\		
&&X_6=\frac{1}{2}
\frac{1}{p^2}\frac{\partial{}}{\partial{p}}
p^{3}
\int_{-1}^{+1}\dee\mu
A(\mu)\Bigg(\int_{-1}^{\mu}\dee\mu
\frac{1}{D_{\mu\mu}}
-\frac{1}{2}\int_{-1}^{+1}\dee\mu
\int_{-1}^{\mu}\dee\mu
\frac{1}{D_{\mu\mu}}\Bigg)
\frac{1}{p^2}\frac{\partial{}}{\partial{p}}
p^{3}
\int_{-1}^{\mu}\dee \mu
A(\mu)g=0,\\		
&&X_7=\frac{1}{2}
\frac{1}{p^2}\frac{\partial{}}{\partial{p}}
p^{3}
\int_{-1}^{+1}\dee\mu
A(\mu)\Bigg(\int_{-1}^{\mu}\dee\mu
\frac{1}{D_{\mu\mu}}
-\frac{1}{2}\int_{-1}^{+1}\dee\mu
\int_{-1}^{\mu}\dee\mu
\frac{1}{D_{\mu\mu}}\Bigg)
\frac{1}{p^2}\frac{\partial{}}{\partial{p}}
p^{4}\frac{\partial{F}}{\partial{p}}	
N\int_{-1}^{\mu}\dee \mu
D_{\mu\mu}
\nonumber\\
&&		
=0,\\	
&&X_8=\frac{1}{2}
\frac{1}{p^2}\frac{\partial{}}{\partial{p}}
p^{3}
\int_{-1}^{+1}\dee\mu
A(\mu)\Bigg(\int_{-1}^{\mu}\dee\mu
\frac{1}{D_{\mu\mu}}
-\frac{1}{2}\int_{-1}^{+1}\dee\mu
\int_{-1}^{\mu}\dee\mu
\frac{1}{D_{\mu\mu}}\Bigg)
N\frac{1}{p^2}\frac{\partial{}}{\partial{p}}
p^{4}\frac{\partial{}}{\partial{p}}
\int_{-1}^{\mu}\dee \mu
D_{\mu\mu}	g=0.	
\end{eqnarray}
With all of the latter formulas,  
we find
\begin{eqnarray}
T_2=-\kappa_{pt}^{(2)}
\frac{1}{p^2}\frac{\partial{}}{\partial{p}}
p^{3}	
\frac{\partial{F}}{\partial{t}}
-\kappa_{pz}^{(2)}
\frac{1}{p^2}\frac{\partial{}}{\partial{p}}
p^{4}
\frac{\partial{F}}{\partial{z}}	
+\kappa_{pp}^{(2)}
\frac{1}{p^2}\frac{\partial{}}{\partial{p}}
p
\frac{\partial{}}{\partial{p}}
p^{3} F		
\end{eqnarray}
with
\begin{eqnarray}
&&\kappa_{pt}^{(2)}=\frac{1}{2}
\int_{-1}^{+1}\dee\mu
A(\mu)\Bigg(\int_{-1}^{\mu}\dee\mu
\frac{1}{D_{\mu\mu}}
-\frac{1}{2}\int_{-1}^{+1}\dee\mu
\int_{-1}^{\mu}\dee\mu
\frac{1}{D_{\mu\mu}}\Bigg)(\mu+1),\\		
&&\kappa_{pz}^{(2)}=	\frac{1}{2}
\int_{-1}^{+1}\dee\mu
A(\mu)\Bigg(\int_{-1}^{\mu}\dee\mu
\frac{1}{D_{\mu\mu}}
-\frac{1}{2}\int_{-1}^{+1}\dee\mu
\int_{-1}^{\mu}\dee\mu
\frac{1}{D_{\mu\mu}}\Bigg)
\frac{\mu^2-1}{2},\\
&&\kappa_{pp}^{(2)}=
\frac{1}{2}
\int_{-1}^{+1}\dee\mu
A(\mu)\Bigg(\int_{-1}^{\mu}\dee\mu
\frac{1}{D_{\mu\mu}}
-\frac{1}{2}\int_{-1}^{+1}\dee\mu
\int_{-1}^{\mu}\dee\mu
\frac{1}{D_{\mu\mu}}\Bigg)
\int_{-1}^{\mu}\dee \mu A(\mu).
\end{eqnarray}

\subsection{$T_3$}

Similarly, we find
\begin{eqnarray}
T_3=\frac{1}{p^2}\frac{\partial{}}{\partial{p}}
p^{4}\frac{\partial{}}{\partial{p}}	
\frac{N}{2}\int_{-1}^{+1}\dee\mu
D_{\mu\mu}	g=0		
\end{eqnarray}

\subsection{The momentum transport equation}

Inserting the formulas for $T_1$, $T_2$, and 
$T_3$ into Equation
(\ref{momentum equation with 3T}) yields
\begin{eqnarray}
\frac{\partial{F}}{\partial{t}}
&=&-\kappa_p
\frac{1}{p^2}\frac{\partial{}}{\partial{p}}
p^{3}F
+
\kappa_{pp}^{(0)}
\frac{1}{p^2}\frac{\partial{}}{\partial{p}}
p^{4}
\frac{\partial{F}}{\partial{p}}	
-\kappa_{zp}^{(1)}	
\frac{1}{p}\frac{\partial{}}{\partial{p}}
p^{3}\frac{\partial{F}}{\partial{z}}
\nonumber\\
&&
-\kappa_{pt}^{(2)}
\frac{1}{p^2}\frac{\partial{}}{\partial{p}}
p^{3}	
\frac{\partial{F}}{\partial{t}}
-\kappa_{pz}^{(2)}
\frac{1}{p^2}\frac{\partial{}}{\partial{p}}
p^{4}
\frac{\partial{F}}{\partial{z}}	
+\kappa_{pp}^{(2)}
\frac{1}{p^2}\frac{\partial{}}{\partial{p}}
p
\frac{\partial{}}{\partial{p}}
p^{3} F
\label{momentum equation with first-order spatial derivative}					
\end{eqnarray}
with
\begin{eqnarray}
&&\kappa_p=	\frac{1}{2}\int_{-1}^{+1}\dee\mu A(\mu),\\
&&\kappa_{pp}^{(0)}=\frac{N}{2}\int_{-1}^{+1}\dee\mu
D_{\mu\mu},\\
&&\kappa_{pt}^{(2)}=\frac{1}{2}
\int_{-1}^{+1}\dee\mu
A(\mu)\Bigg(\int_{-1}^{\mu}\dee\mu
\frac{1}{D_{\mu\mu}}
-\frac{1}{2}\int_{-1}^{+1}\dee\mu
\int_{-1}^{\mu}\dee\mu
\frac{1}{D_{\mu\mu}}\Bigg)(\mu+1),\\
&&\kappa_{pz}^{(2)}=	\frac{1}{2}
\int_{-1}^{+1}\dee\mu
A(\mu)\Bigg(\int_{-1}^{\mu}\dee\mu
\frac{1}{D_{\mu\mu}}
-\frac{1}{2}\int_{-1}^{+1}\dee\mu
\int_{-1}^{\mu}\dee\mu
\frac{1}{D_{\mu\mu}}\Bigg)
\frac{\mu^2-1}{2},\\
&&\kappa_{pp}^{(2)}=
\frac{1}{2}
\int_{-1}^{+1}\dee\mu
A(\mu)\Bigg(\int_{-1}^{\mu}\dee\mu
\frac{1}{D_{\mu\mu}}
-\frac{1}{2}\int_{-1}^{+1}\dee\mu
\int_{-1}^{\mu}\dee\mu
\frac{1}{D_{\mu\mu}}\Bigg)
\int_{-1}^{\mu}\dee \mu A(\mu),\\
&&\kappa_{zp}^{(1)}=\frac{1}{2}
\int_{-1}^{+1}\dee\mu\mu 
\int_{-1}^{\mu}\dee\mu
\frac{1}{D_{\mu\mu}}
\int_{-1}^{\mu}\dee \mu A(\mu).				
\end{eqnarray}
The momentum transport equation
(\ref{momentum equation with first-order 
spatial derivative}) contains terms
involving the first-order 
spatial derivative operator 
$\partial{}/\partial{p}$.
However, this equation
does not contain the spatial convection
term. 
Therefore, we cannot perform 
the iteration for the 
first-order spatial convection. 
For simplicity, 
we neglect the terms in Equation 
(\ref{momentum equation with first-order spatial derivative}) 
involving only the first-order 
spatial derivative operator  
\begin{eqnarray}
\frac{\partial{F}}{\partial{t}}
&=&-\kappa_p
\frac{1}{p^2}\frac{\partial{}}{\partial{p}}
p^{3}F
+
\kappa_{pp}^{(0)}
\frac{1}{p^2}\frac{\partial{}}{\partial{p}}
p^{4}
\frac{\partial{F}}{\partial{p}}	
+\kappa_{pp}^{(2)}
\frac{1}{p^2}\frac{\partial{}}{\partial{p}}
p
\frac{\partial{}}{\partial{p}}
p^{3} F
-\kappa_{pt}^{(2)}
\frac{1}{p^2}\frac{\partial{}}{\partial{p}}
p^{3}	
\frac{\partial{F}}{\partial{t}}	
\label{A42}		
\end{eqnarray}
Inserting Equation 
(ref{A42})
into itself, we obtain
\begin{eqnarray}
\frac{\partial{F}}{\partial{t}}
&=&-\kappa_p
\frac{1}{p^2}\frac{\partial{}}{\partial{p}}
\left(p^{3}F\right)
+
\kappa_{pp}^{(0)}
\frac{1}{p^2}\frac{\partial{}}{\partial{p}}
\left(p^{4}
\frac{\partial{F}}{\partial{p}}\right)	
+\Bigg(\kappa_{pt}^{(2)}\kappa_p
+\kappa_{pp}^{(2)}
\Bigg)
\frac{1}{p^2}\frac{\partial{}}{\partial{p}}
\left[p	
\frac{\partial{}}{\partial{p}}
\left(p^{3}F\right)\right]	
\end{eqnarray}
The latter equation can be rewritten as
\begin{eqnarray}
\frac{\partial{F}}{\partial{t}}
&=&-\mathcal{K}_p
\frac{1}{p^2}\frac{\partial{}}{\partial{p}}
\left(p^{3}F\right)
+
\mathcal{K}_1
\frac{1}{p^2}\frac{\partial{}}{\partial{p}}
\left(p^{4}
\frac{\partial{F}}{\partial{p}}\right)	
+\mathcal{K}_2
\frac{1}{p^2}\frac{\partial{}}{\partial{p}}
\left[p	
\frac{\partial{}}{\partial{p}}
\left(p^{3}F\right)\right]		
\end{eqnarray}
with
\begin{eqnarray}
&&\mathcal{K}_p=\kappa_p,\\
&&\mathcal{K}_1=\kappa_{pp}^{(0)},\\
&&\mathcal{K}_2=\kappa_{pt}^{(2)}\kappa_p
+\kappa_{pp}^{(2)}.
\end{eqnarray}

\section{The derivation  of the 
formula for the moment of the fractional transport 
equation}
\label{The derivation  of moment
formula for the fractional transport 
equation with the formulas $I_1$ and $I_2$}	

After integrating by parts, we can rewrite
the integrals $I_1$ and $I_2$ as
\begin{eqnarray}
	&&
	I_1=J_1+J_2+J_3,
	\\
	&&
	I_2=J_4+J_5+J_6
\end{eqnarray}
with
\begin{eqnarray}
	&&
	J_1=
	\int_{-\infty}^{0} dz
	|z|^{\alpha-n}
	\int_{-\infty}^z d\xi\left(z-\xi\right)
	^{n-\beta-1}F(\xi),
	\\
	&&
	J_2
	=
	\int_{0}^{+\infty} dz
	|z|^{\alpha-n}
	\int_{-\infty}^0 d\xi\left(z-\xi\right)
	^{n-\beta-1}F(\xi),
	\\
	&&
	J_3
	=
	\int_{0}^{+\infty} dz
	|z|^{\alpha-n}
	\int_{0}^z d\xi\left(z-\xi\right)
	^{n-\beta-1}F(\xi),
	\\
	&&
	J_4
	=
	\int_{-\infty}^{0} dz
	|z|^{\alpha-n}
	\int_{z}^0 d\xi\left(z-\xi\right)
	^{n-\beta-1}F(\xi),
	\\
	&&
	J_5
	=
	\int_{-\infty}^{0} dz
	|z|^{\alpha-n}
	\int_{0}^{+\infty} d\xi\left(z-\xi\right)
	^{n-\beta-1}F(\xi),
	\\
	&&
	J_6
	=
	\int_{0}^{+\infty} dz
	|z|^{\alpha-n}
	\int_{z}^{+\infty} d\xi\left(z-\xi\right)
	^{n-\beta-1}F(\xi).
\end{eqnarray}

Defining $\dee z=\xi\dee\eta$,
we can rewrite $J_1$ as
\begin{eqnarray}
	J_1
	&=&
	\frac{1}{(-1)^{\beta-n}}
	\int_{-\infty}^{0} d\xi F(\xi)
	|\xi|^{\alpha-n}
	\int_{0}^1 d\eta \eta^{\alpha-n}
	(1-\eta)^{n-\beta-1}
	\left(z-\xi\right)
	^{n-\beta-1}
	\nonumber\\
	&=&
	\frac{1}{(-1)^{\beta-n}}
	\frac{\Gamma(\alpha-n+1)\Gamma(n-\beta)}
	{\Gamma(\alpha-\beta+1)}
	\int_{-\infty}^0
	\dee \xi
	F(\xi) |\xi|^{\alpha-\beta}.
\end{eqnarray}
Similarly, with $\xi=-\xi'$ and $z=-z'$, we 
obtain
\begin{eqnarray}
	J_2
	=
	\int_{0}^{+\infty} d\xi' F(-\xi')
	\int_{-\infty}^0 dz'
	\left(\xi'-z'\right)
	^{n-\beta-1}
	|z'|^{\alpha-n}.
\end{eqnarray}
In addition, using the relation $z=\xi/\eta$, 
we derive
\begin{eqnarray}
	&&J_3
	=
	\frac{\Gamma(\beta-\alpha)\Gamma(n-\beta)}
	{\Gamma(n-\alpha)}
	\int_0^{+\infty}
	\dee \xi
	F(\xi) |\xi|^{\alpha-\beta},\\
	&&J_4
	=
	\frac{1}{(-1)^{\alpha-n}}
	\frac{\Gamma(\beta-\alpha)\Gamma(n-\beta)}
	{\Gamma(n-\alpha)}
	\int_{-\infty}^0
	\dee \xi
	F(\xi) |\xi|^{\alpha-\beta},\\
	&&
	J_5
	=
	\int_{0}^{+\infty} d\xi F(\xi)
	\int_{-\infty}^0 dz
	\left(\xi-z\right)^{n-\beta-1}
	|z|^{\beta-n}.
\end{eqnarray}
With $z=\xi\eta$, $J_6$ becomes
\begin{eqnarray}
	J_6
	=
	\frac{\Gamma(\alpha-n+1)\Gamma(n-\beta)}
	{\Gamma(\alpha-\beta+1)}
	\int_0^{+\infty}
	\dee \xi
	F(\xi) |\xi|^{\alpha-\beta}.
\end{eqnarray}
Thus, Equation
(\ref{D_t^gamma z^alpha=C(I1+I2)})
becomes
\begin{eqnarray}
	\int_{-\infty}^{+\infty}\dee z |z|^{\alpha}
	\frac{\partial{F}}{\partial{t}}
	=
	(-1)^{n+1}\frac{\chi}{2\cos \frac{\pi\beta}{2}}
	\frac{\Gamma(\alpha+1)}{\Gamma(n-\beta)
		\Gamma(\alpha-n+1)} 
	W.
\label{Appendix-fractional equation 100}
\end{eqnarray}
Here,
\begin{eqnarray}
	W
	=W_1+W_2+W_3
\end{eqnarray}
with
\begin{eqnarray}
	&&
	W_1
	=
	J_1+(-1)^n J_6,
	\\
	&&
	W_2
	=
	J_2+(-1)^n J_5,
	\\
	&&
	W_3
	=
	J_3+(-1)^n J_4. 
\end{eqnarray}

If we assume the distribution
function is an even function in
$x$, i.e., $f(\xi)=f(-\xi)$,
we derive
\begin{eqnarray}
	W_1&=&\frac{1}{(-1)^{\beta-n}}
	\frac{\Gamma(\alpha-n+1)\Gamma(n-\beta)}
	{\Gamma(\alpha-\beta+1)}
	\int_{-\infty}^0
	\dee \xi
	F(\xi) |\xi|^{\alpha-\beta} 
	+
	(-1)^n 
	\frac{\Gamma(\alpha-n+1)\Gamma(n-\beta)}
	{\Gamma(\alpha-\beta+1)}
	\int_0^{+\infty}
	\dee \xi
	F(\xi) |\xi|^{\alpha-\beta}
	\nonumber\\
	&& 
	=\frac{1}{2}
	(-1)^n 
	\frac{\Gamma(\alpha-n+1)\Gamma(n-\beta)}
	{\Gamma(\alpha-\beta+1)}
	\left[
	\frac{1}{(-1)^{\beta}}
	+
	1
	\right]
	\left\langle 
	|z|^{\alpha-\beta}
	\right\rangle
\end{eqnarray}
and
\begin{eqnarray}
	W_2
	=
	\left[
	1+(-1)^n
	\right]
	\frac{1}{2}
	\int_{-\infty}^0\dee\xi F(\xi)
	\int_0^{+\infty}\dee z
	\frac{|z|^{\beta-n}}
	{(z-\xi)^{\alpha-n+1}}
\end{eqnarray}
Similarly, for $F(\xi)=F(-\xi)$ we have
\begin{eqnarray}
	W_3
	=
	\frac{1}{2}
	\frac{\Gamma(\beta-\alpha)\Gamma(n-\beta)}
	{\Gamma(n-\alpha)}
	\left[
	1
	+
	\frac{1}{(-1)^{\alpha}}
	\right]
	\left\langle 
	|\xi|^{\alpha-\beta}
	\right\rangle
\end{eqnarray}
and
\begin{eqnarray}
	W
	&=&	
	\frac{1}{2}
	(-1)^n 
	\frac{\Gamma(\alpha-n+1)\Gamma(n-\beta)}
	{\Gamma(\alpha-\beta+1)}
	\left[
	\frac{1}{(-1)^{\beta}}
	+
	1
	\right]
	\left\langle 
	|\xi|^{\alpha-\beta}
	\right\rangle
	\nonumber\\
	&&
	+
	\frac{1}{2}
	\frac{\Gamma(\beta-\alpha)\Gamma(n-\beta)}
	{\Gamma(n-\alpha)}
	\left[
	1
	+
	\frac{1}{(-1)^{\alpha}}
	\right]
	\left\langle 
	|\xi|^{\alpha-\beta}
	\right\rangle
\end{eqnarray}

Since $0<\beta<\alpha<n=1$ and $\beta<1/2$,
\begin{eqnarray}
	W&=&
	\frac{1}{2}
	\Bigg\{
	\frac{\Gamma(\beta-\alpha)}
	{\Gamma(n-\alpha)}
	\left(
	1
	+
	\frac{1}{(-1)^{\alpha}}
	\right)
	-
	\frac{\Gamma(\alpha-n+1)}
	{\Gamma(\alpha-\beta+1)}
	\left(
	1+
	\frac{1}{(-1)^{\beta}}
	\right)
	\Bigg\}
	\Gamma(n-\beta)
	\left\langle 
	|\xi|^{\alpha-\beta}
	\right\rangle
	\nonumber\\
	&=&
	\Bigg(
	\frac{\Gamma(\beta-\alpha)}
	{\Gamma(n-\alpha)}
	\cos^2\frac{\alpha\pi}{2}
	-
	\frac{\Gamma(\alpha-n+1)}
	{\Gamma(\alpha-\beta+1)}
	\cos^2\frac{\beta\pi}{2}
	\Bigg)
	\Gamma(n-\beta)
	\left\langle 
	|z|^{\alpha-\beta}
	\right\rangle.
\end{eqnarray}
Inserting the latter equation into 
Equation
(\ref{Appendix-fractional equation 100})
gives
\begin{eqnarray}
\frac{\dee}{\dee t}
\left\langle
|z|^{\alpha}
\right\rangle
=&&\chi
\frac{(-1)^{n+1}}{2\cos \frac{\pi\beta}{2}}
\frac{\Gamma(\alpha+1)}{\Gamma(n-\beta)
\Gamma(\alpha-n+1)} 
\Bigg(
\frac{\Gamma(\beta-\alpha)}
{\Gamma(n-\alpha)}
\cos^2\frac{\alpha\pi}{2}
\nonumber\\
&&
-
\frac{\Gamma(\alpha-n+1)}
{\Gamma(\alpha-\beta+1)}
\cos^2\frac{\beta\pi}{2}
\Bigg)
\Gamma(n-\beta)
\left\langle 
|z|^{\alpha-\beta}
\right\rangle,
\end{eqnarray}
from which the formula of coefficient $\chi$
can be obtained. 
	
\end{appendices}

\end{CJK*}

\end{document}